\documentclass[useAMS,usenatbib]{mn2e}

\usepackage{graphicx}

\newcommand{\LO}{\ensuremath{L_{\odot}}}   
\newcommand{\MO}{\ensuremath{M_{\odot}}}   

\newcommand{\aap}{A\&A}     
\newcommand{\aj}{AJ}    
\newcommand{\apj}{ApJ}     
\newcommand{\apjl}{ApJL}   
\newcommand{\apjs}{ApJS}   
\newcommand{\araa}{ARA\&A}  
\newcommand{\mnras}{MNRAS}     
\newcommand{\nat}{Nature}  
\newcommand{\pasj}{PASJ}        
\newcommand{\pasp}{PASP}   
\newcommand{\physrep}{Phys.~Rep.} 
\newcommand{\procspie}{Proc.~SPIE}

\let\lsim=\la
\let\gsim=\ga

\title
[AzTEC/ASTE 1.1~mm Survey of the ADF-S]
{AzTEC/ASTE 1.1~mm Survey of the AKARI Deep Field South: Source Catalogue and Number Counts}
\author[B. Hatsukade et al.]
		{B. Hatsukade,$^{1}$\thanks{E-mail:hatsukade@nro.nao.ac.jp}
 		K. Kohno,$^{2,3}$
 		I. Aretxaga,$^{4}$
 		J. E. Austermann,$^{5}$
 		H. Ezawa,$^{6}$
 		\newauthor
 		D. H. Hughes,$^{4}$
 		S. Ikarashi,$^{2}$
 		D. Iono,$^{1}$
 		R. Kawabe,$^{1}$ 
 		S. Khan,$^{7}$
 		H. Matsuo,$^{6}$
 		\newauthor
 		S. Matsuura,$^{8}$
 		K. Nakanishi,$^{6}$
 		T. Oshima,$^{1}$
 		T. Perera,$^{9}$
 		K. S. Scott,$^{10}$
 		M. Shirahata,$^{8}$
 		\newauthor
 		T. T. Takeuchi,$^{11}$
 		Y. Tamura,$^{1}$
 		K. Tanaka,$^{12}$
 		T. Tosaki,$^{13}$
 		G. W. Wilson,$^{9}$
 		M. S. Yun$^{9}$ \\
$^{1}$Nobeyama Radio Observatory, Minamimaki, Minamisaku, Nagano 384-1805, Japan \\
$^{2}$Institute of Astronomy, the University of Tokyo, 2-21-1 Osawa, Mitaka, Tokyo 181-0015, Japan \\
$^{3}$Research Center for the Early Universe, University of Tokyo, 7-3-1 Hongo, Bunkyo, Tokyo 113-0033, Japan \\
$^{4}$Instituto Nacional de Astrofisica, \'{O}ptica y Electr\'{o}nica (INAOE), Aptdo. Postal 51 y 216, 72000 Puebla, Pue., Mexico \\
$^{5}$Center for Astrophysics and Space Astronomy, University of Colorado, Boulder, CO 80309, USA \\
$^{6}$National Astronomical Observatory of Japan, 2-21-1 Osawa, Mitaka, Tokyo 181-8588, Japan \\
$^{7}$Shanghai Key Lab for Astrophysics, Shanghai Normal University, Shanghai 200234, China \\
$^{8}$Institute of Space and Astronautical Science, Japan Aerospace Exploration Agency, 3-1-1 Yoshinodai, Sagamihara, Kanagawa 229-8510, Japan \\
$^{9}$University of Massachusetts, Department of Astronomy, Amherst, MA01003, USA \\
$^{10}$Department of Physics and Astronomy, University of Pennsylvania, Philadelphia, PA 19104, USA \\
$^{11}$Institute for Advanced Research, Nagoya University, Chikusa-ku, Nagoya 464-8601, Japan \\
$^{12}$Department of Physics, Faculty of Science and Technology, Keio University, 3-14-1 Hiyoshi, Kohoku-ku, Yokohama, Kanagawa 223-8522, Japan \\
$^{13}$Joetsu University of Education, Yamayashiki, Joetsu, Niigata 943-8512, Japan \\
}

\begin{document}

\date{Accepted 2010 September 7. Received 2010 August 18; in original form 2010 June 21}

\pagerange{\pageref{firstpage}--\pageref{lastpage}} \pubyear{2010}

\maketitle

\label{firstpage}

\begin{abstract}
We present results of a 1.1~mm deep survey of the AKARI Deep Field South (ADF-S) with AzTEC mounted on the Atacama Submillimetre Telescope Experiment (ASTE). 
We obtained a map of 0.25~deg$^2$ area with an rms noise level of 0.32--0.71~mJy. 
This is one of the deepest and widest maps thus far at millimetre and submillimetre wavelengths. 
We uncovered 198 sources with a significance of 3.5--15.6$\sigma$, providing the largest catalog of 1.1~mm sources in a contiguous region. 
Most of the sources are not detected in the far-infrared bands of the {\sl AKARI} satellite, suggesting that they are mostly at $z \ge 1.5$ given the detection limits. 
We constructed differential and cumulative number counts in the ADF-S, the Subaru/{\sl XMM Newton} Deep Field (SXDF), and the SSA~22 field surveyed by AzTEC/ASTE, which provide currently the tightest constraints on the faint end. 
The integration of the best-fit number counts in the ADF-S find that the contribution of 1.1~mm sources with fluxes $\ge$1~mJy to the cosmic infrared background (CIB) at 1.1~mm is 12--16\%, suggesting that the large fraction of the CIB originates from faint sources of which the number counts are not yet constrained. 
We estimate the cosmic star-formation rate density contributed by 1.1~mm sources with $\ge$1~mJy using the best-fit number counts in the ADF-S and find that it is lower by about a factor of 5--10 compared to those derived from UV/optically-selected galaxies at $z \sim 2$--3. 
The fraction of stellar mass of the present-day universe produced by 1.1~mm sources with $\ge$1~mJy at $z \ge 1$ is $\sim$20\%, calculated by the time integration of the star-formation rate density. 
If we consider the recycled fraction of $>$0.4, which is the fraction of materials forming stars returned to the interstellar medium, the fraction of stellar mass produced by 1.1~mm sources decrease to $\lsim$10\%.

\end{abstract}

\section{Introduction}

Over the past decade, millimetre and submillimetre observations have shown that (sub)millimetre-bright galaxies (hereafter SMGs) hold important clues to galaxy evolution and the cosmic star formation history \cite[for a review]{blai02}. 
SMGs are highly obscured by dust, and the resulting thermal dust emission dominates the bolometric luminosity. 
The source of heating energy is dominated by vigorous star formation with star formation rates (SFRs) of several 100--1000 $\MO\ \rm{yr}^{-1}$. 
Optical/near-infrared spectroscopy of a sample of SMGs with radio counterparts revealed a median redshift of $z \sim 2$ for the population \citep{swin04, chap05}. 
Recently, SMGs at $z > 4$ have been confirmed \citep{capa08, copp09, dadd09, knud10}  and there is now a spectroscopically confirmed source at $z=5.3$ \citep{riec10}.

Coupled with reports of high dynamical mass and gas mass \citep[e.g.,][]{grev05, tacc06}, it is suggested that SMGs are the progenitors of massive spheroidal galaxies observed during their formation phase \citep[e.g.,][]{lill96, smai04}.

Mounting evidence shows that the cosmic infrared background \citep[CIB][]{puge96, fixs98} at millimetre and submillimetre wavelengths is largely contributed by high-redshift galaxies \citep{laga05}. 
The CIB is the integral of unresolved emission from extragalactic sources and contains information on the evolutionary history of galaxies. 
While 850~$\mu$m surveys have resolved 20--40\% of the CIB into point sources in blank fields \citep[e.g.,][]{eale00, bory03, copp06} and 50--100\% in lensing cluster fields \citep[e.g,][]{blai02, cowi02, knud08}, 1~mm blank field surveys have resolved only $\sim$10\% \citep[e.g.,][]{grev04, laur05, malo05, scot08, pere08, scot10}. 
A large portion of the CIB at millimetre and submillimetre wavelengths likely arises from galaxies with fainter flux densities. 

In conjunction with constraints from the CIB, the number counts of SMGs are sensitive to the history of galaxy evolution at high redshifts. 
This requires constraining both the faint and bright end of the number counts, which in turn requires a suitable combination of small, deep surveys along with shallower wide-area surveys. 
Blank field surveys at millimetre and submillimetre wavelengths have been carried out with large bolometer arrays such as the Submillimetre Common User Bolometer Array \citep[SCUBA;][]{holl99} on the James Clerk Maxwell Telescope (JCMT) \citep[e.g.,][]{smai97, hugh98, copp06}, the Max-Plank Millimetre Bolometer Array \citep[MAMBO;][]{krey98} on the IRAM 30-m telescope and the Bolocam \citep{glen98} on the Caltech Submillimetre Observatory (CSO) \citep[e.g.,][]{grev04, grev08, bert07, laur05}, 
the Large Apex BOlometer CAmera \citep[LABOCA;][]{siri09} on the Atacama Pathfinder EXperiment (APEX) \citep[e.g.,][]{weis09, swin10}, and AzTEC \citep{wils08a} on the JCMT \citep{scot08, pere08, aust09, aust10}. 
However, the total area covered in existing surveys is still small ($\lsim$1~deg$^2$) compared to the cosmic large scale structure, and substantial field-to-field variations can be seen in the published number counts. 
In addition, because of the limited depth of these surveys, the number counts of SMGs at faint flux densities ($\sim$1~mJy) are still not well constrained.

We performed extensive surveys at 1.1~mm with AzTEC mounted on the Atacama Submillimetre Telescope Experiment \citep[ASTE;][]{ezaw04, ezaw08} in 2007 and 2008 \citep{wils08b, tamu09, scot10}. 
The ASTE is a 10-m submillimetre telescope located at Pampa la Bola in the Atacama desert in Chile. 
Some of the AzTEC/ASTE sources are followed up by submm/mm interferometers \citep[][Tamura et al. in preparation]{hats10, ikar10}. 
In this paper, we report on a deep blank field survey of the AKARI Deep Field-South (ADF-S). 
The ADF-S is a multi-wavelength deep survey field near the South Ecliptic Pole. 
It is known to be one of the lowest-cirrus region in the whole sky \citep[100 $\mu$m flux density of $<$0.5~MJy~sr$^{-1}$ and HI column density of $\sim$$5 \times 10^{19}$~cm$^{-2}$;][]{schl98}, providing a window to the high-redshift dusty universe. 
{\sl AKARI}, an infrared satellite \citep{mats05}, has conducted deep surveys with the InfraRed Camera \citep[IRC;][]{onak07} at 2.4, 3.2, 4.1, 7, 11, 15, 24 $\mu$m, and with the Far-Infrared Surveyor \citep[FIS;][]{kawa07} at 65, 90, 140, and 160 $\mu$m, down to the confusion limit \citep{shir09}. 
Multi-wavelength follow-up observations from the UV to the radio are under way.
The full data sets of IR to submillimetre bands with {\sl AKARI}, {\sl Spitzer} \citep{clem10}, the Balloon-borne Large-Aperture Submillimetre Telescope (BLAST), {\sl Herschel Space Observatory}, and the AzTEC/ASTE offers a unique opportunity to study the dusty galaxy population that contributes to the cosmic background at IR--mm wavelengths.

This paper presents the 1.1-mm map and source catalog of the ADF-S. 
Together with the results from the Subaru/{\sl XMM Newton} Deep Field (SXDF; Ikarashi et al. in preparation) and the SSA~22 field \citep{tamu09} surveyed by AzTEC/ASTE, we present statistical properties of the SMG population. 
Currently, these are the deepest wide-area survey ever made at millimetre wavelengths along with the AzTEC/ASTE GOODS-S survey \citep{scot10}, providing the tightest constraints on number counts toward the faint flux density, albeit with lower resolution than has been typically employed to date. 
Comparisons with multi-wavelength data and statistical studies such as clustering analysis of this dataset will be presented in future papers.

The arrangement of this paper is as follows. 
Section~\ref{sec:observations} summarises the observations of the ADF-S with AzTEC/ASTE. 
Section~\ref{sec:reduction} outlines the data reduction and calibration details. 
In Section~\ref{sec:map-catalog} we present the 1.1~mm map and the source catalogue. 
In Section~\ref{sec:counts} we derive number counts of the ADF-S, the SXDF, and the SSA~22 field, and compare them with other 1-mm wavelength surveys and luminosity evolution models. 
In Section~\ref{sec:cib} we estimate the contribution of 1.1~mm sources to the CIB. 
In Section~\ref{sec:comparison} we constrain the redshifts of the AzTEC sources using flux ratios between 1.1~mm and 90~$\mu$m obtained with {\sl AKARI}/FIS. 
In Section~\ref{sec:sfh} we discuss the cosmic star formation history traced by 1.1~mm sources. 
A summary is presented in Section~\ref{sec:summary}.

Throughout the paper, we adopt a cosmology with $H_0=70$ km s$^{-1}$ Mpc$^{-1}$, $h=0.7$, $\Omega_{\rm{M}}=0.3$, and $\Omega_{\Lambda}=0.7$.

\begin{figure*}
\begin{center}
\includegraphics[width=\linewidth]{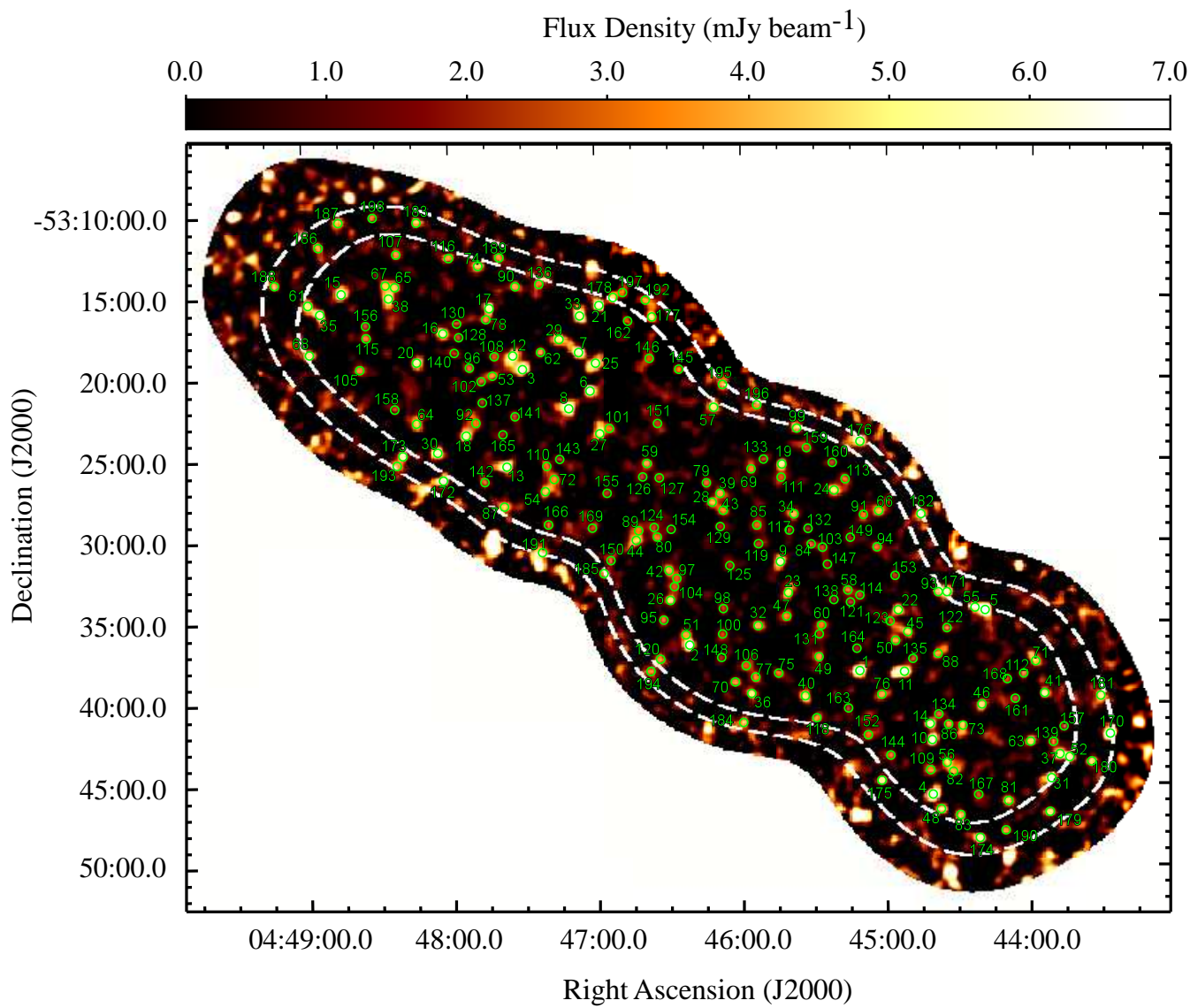}
\end{center}
\caption[Signal map of ADF-S]
{
Signal map of the ADF-S. 
White contours represent the 30\% (outer contour) and 50\% (inner contour) coverage regions (corresponding to rms noise levels of 0.55 and 0.71~mJy~beam$^{-1}$), respectively. 
The 198 sources detected with $\ge$3.5$\sigma$ in the 30\% coverage region are indicated by green circles with a $30''$ diameter, and numbered in order of significance (Table~\ref{tab:catalog}). 
The pixel size is $3'' \times 3''$. 
}
\label{fig:signal-map}
\end{figure*}

\section{Observations}\label{sec:observations}

The center region of the ADF-S was observed with AzTEC on the ASTE. 
The observations were made from September 16 to October 14 in 2007, and from August 4 to December 21 in 2008. 
The operations were carried out remotely from the ASTE operation rooms in San Pedro de Atacama, Chile, and in Mitaka, Japan through the network observation system N-COSMOS3 developed by the National Astronomical Observatory of Japan (NAOJ) \citep{kama2005}. 
The full width at half maximum (FWHM) of the AzTEC detectors on the ASTE is $\sim$$30''$ at 1.1~mm (270 GHz), and the field of view of the array is roughly circular with a diameter of 8$'$. 
During the observing run, 107 and 117 out of the 144 AzTEC detectors were operational in 2007 and 2008, respectively.

In order to maximize the observing efficiency, we used a lissajous scan pattern to map the field. 
We chose a high maximum velocity of $300''$~s$^{-1}$ to mitigate low-frequency atmospheric fluctuations. 
The lissajous scan pattern provides $\sim$$20' \times 20'$ coverage on the sky, in which the central $\sim$$12' \times 12'$ area is nearly uniform, with decreasing integration time from the inside to the outside. 
We covered the ADF-S with seven different field centers to make the noise level of the entire map uniform. 
We obtained a total of 319 individual observations under conditions where the atmospheric zenith opacity at 220 GHz as monitored with a radiometer at the ASTE telescope site was $\tau_{220\ \rm{GHz}}$ = 0.02--0.1. 
The total time spent on-field was $\sim$216 hours. 
Uranus or Neptune was observed at least once a night in raster-scan mode to measure each detector's point spread function (PSF) and relative position, and to determine the flux conversion factor (FCF) for absolute calibration \citep{wils08a}. 
Pointing observations with the quasar J0455-462 that lies $\sim$7 degree from the field center were performed every two hours, bracketing science observations. 
A pointing model for each observing run in 2007 and in 2008 is constructed from these data, and we make corrections to the telescope astrometry. 
The resulting pointing accuracy is better than $3''$ \citep{wils08b}.

\section{Data Reduction}\label{sec:reduction}

The data were reduced in a manner identical to that described in \cite{scot10}. 
We used a principle component analysis (PCA) to remove the low-frequency atmospheric signal from the time-stream data \citep{laur05, scot08}. 
The PCA method AC couples the bolometer time-stream data, making the entire map and point-source kernel have mean of zero. 
The cleaned time-series data is projected into map space using $3'' \times 3''$ pixels in R.A.-Dec., and the individual observations are co-added into a single map by weighted averaging. 
We also create 100 noise realizations by `jackknifing' the time-series data as described in \cite{scot08}. 
These noise maps represent realizations of the underlying random noise in the map in the absence of sources (both bright and confused) and are used throughout this paper to characterize the properties of the map and source catalogue. 
The point source profile is affected by the PCA method since the faint point sources with low spatial frequencies are also attenuated. 
The PCA method makes the mean of the map zero, causing negative side lobes around the peak of the point source profile. 
We trace the effects of PCA and other process in the analysis on the point source profile; this `point source kernel' is used to optimally filter the coadded map and the 100 noise realizations for the detection of point sources. 
A 2-dimensional Gaussian fitting to the point source kernel gives a FWHM of $35\farcs9$.

\section{Map and Source Catalogue}\label{sec:map-catalog}

\subsection{Map}\label{sec:map}
\begin{figure}
\begin{center}
\includegraphics[width=\linewidth]{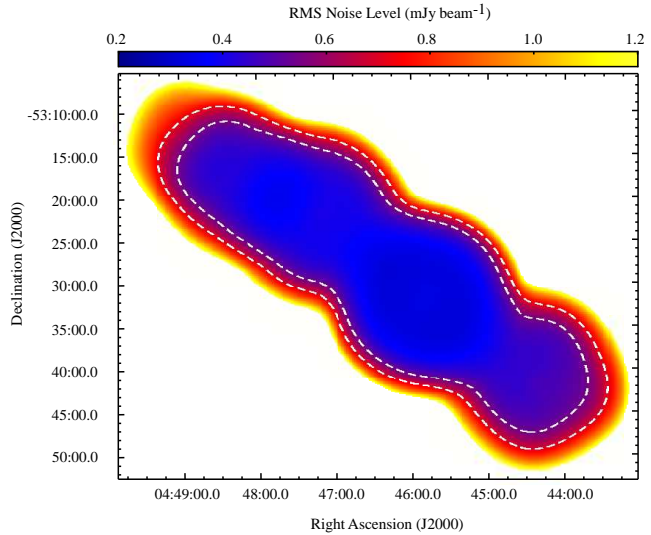}
\end{center}
\caption[Noise map of AzTEC/ASTE ADF-S]
{
Noise map of the AzTEC/ASTE ADF-S. 
The outer and inner contours indicate the 30\% and 50\% coverage region, respectively. 
}
\label{fig:noise-map}
\end{figure}

The signal map and the corresponding noise map are shown in Figures~\ref{fig:signal-map} and \ref{fig:noise-map}, respectively. 
The dashed curves represent regions with $\ge$30\% (outer contour) and $\ge$50\% (inner contour) of the maximum weights (hereafter called as 30\% and 50\% coverage region). 
The area and noise level are 709 and 200~arcmin$^2$, and 0.32--0.55, 0.55--0.71~mJy in the 50\% and 30--50\% coverage region, respectively (Table \ref{tab:coverage}). 
This survey is confusion-limited, where the 5$\sigma$-confusion limit estimated by \cite{take04} is 4.4~mJy using the point source kernel and the differential number counts in the ADF-S derived in \S~\ref{sec:counts_adfs}.

\begin{figure}
\begin{center}
\includegraphics[width=\linewidth]{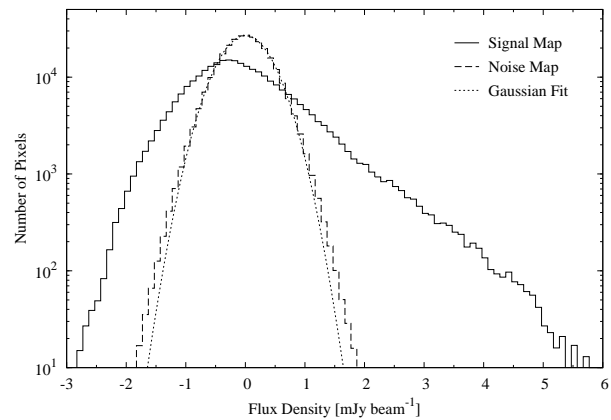}
\end{center}
\caption[Flux distributions of real map and noise maps. ]{
Distribution of flux density in the real map (solid histogram) and averaged over the 100 noise realizations (dashed histogram) in the 50\% coverage region. 
The result of a Gaussian fit to the flux distribution of the noise maps is plotted as a dashed curve. 
}
\label{fig:flux_hist}
\end{figure}

Fig. \ref{fig:flux_hist} shows the distribution of flux values in the map, compared to that averaged over the 100 noise realizations within the 50\% coverage region. 
The result of a Gaussian fit to the averaged noise map is superimposed in Fig. \ref{fig:flux_hist}.  
The presence of real sources in the map makes excess of both positive and negative valued pixels over the histogram of the noise map, since the signal map is created to have a mean of zero. 
This fit deviates from the distribution of pixel values at high positive and negative fluxes because the map is not uniform over the entire region, with the outer region being slightly noisier.

\begin{table}
\centering
\begin{minipage}{\linewidth}
\caption{
Map properties in the 50\% and the 30--50\% coverage regions.
}
\label{tab:coverage}
\begin{tabular}{@{}ccccc@{}}
\hline
Coverage & 
Area & 
Noise level & 
Source\footnote{Number of sources ($\ge$3.5$\sigma$).} & 
False Detections\footnote{Number of false detection ($\ge$3.5$\sigma$).} \\ 
(\%) & (arcmin$^2$) & (mJy beam$^{-1}$) & & \\
\hline
50\%     & 709 & 0.32--0.55 & 169 & $4.9 \pm 0.22$ \\
30--50\% & 200 & 0.55--0.71 &  29 & $1.4 \pm 0.12$ \\
\hline
\end{tabular}
\end{minipage}
\end{table}

\subsection{Source Catalogue}\label{sec:catalog}

Source extraction is performed on the SN map using a criterion of $\ge$3.5$\sigma$. 
The source positions are determined by flux-squared weighting on pixels within a $15''$ radius of the nominal peak. 
We detect 198 and 169 sources with a significance of 3.5--15.6$\sigma$ in the 30\% and 50\% coverage regions, respectively. 
The source catalog is given in Table \ref{tab:catalog}, where both the observed and the deboosted flux densities (\S~\ref{sec:deboosting}) are listed. 
169 sources (ADFS-AzTEC1--169) detected within the 50\% coverage region, the deeper and more uniform coverage region, are listed first, followed by the remaining 29 sources (ADFS-AzTEC170--198) detected outside the 50\% coverage region.

\begin{table*}
\caption{
AzTEC/ASTE ADF-S 3.5$\sigma$ Source Catalog. 
AzTEC1--169 are detected in the 50\% coverage region and AzTEC170--198 are detected in the region with coverage 30--50\%. 
The columns give 
(1) Source name; 
(2) ID; 
(3) Right ascension; 
(4) Declination; 
(5) Observed flux density and 1$\sigma$ error; 
(6) Deboosted flux density and 68\% confidence level; 
(7) Signal to noise ratio. 
}
\label{tab:catalog}
\centering
\begin{tabular}{@{}clccccc@{}}
\hline
Name & 
\multicolumn{1}{c}{ID} & 
R.A. & 
Dec. & 
$S_{\rm observed}$ &
$S_{\rm{deboosted}}$ &
S/N \\
& & ($^h$ $^m$ $^s$) & ($^{\circ}$ $'$ $''$) & (mJy) & (mJy) & \\
\hline
AzTEC J044511.52$-$533807.3& ADFS-AzTEC1   & 4 45 11.52 & $-53$ 38 07.3 & $5.9 \pm 0.4$ & $5.7^{+0.4}_{-0.4}$ & 15.6 \\
AzTEC J044621.97$-$533630.4& ADFS-AzTEC2   & 4 46 21.97 & $-53$ 36 30.4 & $5.5 \pm 0.4$ & $5.3^{+0.5}_{-0.4}$ & 14.2 \\
AzTEC J044730.07$-$531928.1& ADFS-AzTEC3   & 4 47 30.07 & $-53$ 19 28.1 & $5.1 \pm 0.4$ & $4.9^{+0.5}_{-0.4}$ & 13.8 \\
AzTEC J044441.16$-$534543.4& ADFS-AzTEC4   & 4 44 41.16 & $-53$ 45 43.4 & $6.5 \pm 0.5$ & $6.2^{+0.5}_{-0.6}$ & 12.8 \\
AzTEC J044419.58$-$533422.9& ADFS-AzTEC5   & 4 44 19.58 & $-53$ 34 22.9 & $6.7 \pm 0.5$ & $6.4^{+0.6}_{-0.6}$ & 12.2 \\
AzTEC J044702.37$-$532049.8& ADFS-AzTEC6   & 4 47 02.37 & $-53$ 20 49.8 & $4.9 \pm 0.4$ & $4.7^{+0.4}_{-0.5}$ & 12.1 \\
AzTEC J044706.89$-$531827.3& ADFS-AzTEC7   & 4 47 06.89 & $-53$ 18 27.3 & $4.9 \pm 0.4$ & $4.7^{+0.4}_{-0.5}$ & 12.0 \\
AzTEC J044711.18$-$532154.4& ADFS-AzTEC8   & 4 47 11.18 & $-53$ 21 54.4 & $4.9 \pm 0.4$ & $4.7^{+0.5}_{-0.5}$ & 12.0 \\
AzTEC J044544.27$-$533124.0& ADFS-AzTEC9   & 4 45 44.27 & $-53$ 31 24.0 & $4.0 \pm 0.3$ & $3.8^{+0.4}_{-0.4}$ & 11.9 \\
AzTEC J044441.47$-$534222.3& ADFS-AzTEC10  & 4 44 41.47 & $-53$ 42 22.3 & $5.3 \pm 0.5$ & $5.1^{+0.5}_{-0.5}$ & 11.8 \\
AzTEC J044452.99$-$533810.5& ADFS-AzTEC11  & 4 44 52.99 & $-53$ 38 10.5 & $4.8 \pm 0.4$ & $4.5^{+0.5}_{-0.5}$ & 11.4 \\
AzTEC J044733.99$-$531837.5& ADFS-AzTEC12  & 4 47 33.99 & $-53$ 18 37.5 & $4.0 \pm 0.4$ & $3.8^{+0.4}_{-0.4}$ & 11.0 \\
AzTEC J044736.90$-$532527.5& ADFS-AzTEC13  & 4 47 36.90 & $-53$ 25 27.5 & $4.5 \pm 0.4$ & $4.3^{+0.5}_{-0.5}$ & 10.7 \\
AzTEC J044442.42$-$534122.9& ADFS-AzTEC14  & 4 44 42.42 & $-53$ 41 22.9 & $4.8 \pm 0.4$ & $4.5^{+0.5}_{-0.5}$ & 10.6 \\
AzTEC J044844.20$-$531443.4& ADFS-AzTEC15  & 4 48 44.20 & $-53$ 14 43.4 & $5.0 \pm 0.5$ & $4.7^{+0.5}_{-0.5}$ & 10.5 \\
AzTEC J044802.69$-$531712.7& ADFS-AzTEC16  & 4 48 02.69 & $-53$ 17 12.7 & $4.2 \pm 0.4$ & $3.9^{+0.4}_{-0.5}$ & 10.5 \\
AzTEC J044743.53$-$531541.8& ADFS-AzTEC17  & 4 47 43.53 & $-53$ 15 41.8 & $4.2 \pm 0.4$ & $4.0^{+0.5}_{-0.5}$ & 10.3 \\
AzTEC J044753.52$-$532331.4& ADFS-AzTEC18  & 4 47 53.52 & $-53$ 23 31.4 & $4.1 \pm 0.4$ & $3.9^{+0.5}_{-0.4}$ & 10.3 \\
AzTEC J044543.44$-$532524.3& ADFS-AzTEC19  & 4 45 43.44 & $-53$ 25 24.3 & $3.8 \pm 0.4$ & $3.6^{+0.4}_{-0.4}$ & 10.1 \\
AzTEC J044813.58$-$531859.6& ADFS-AzTEC20  & 4 48 13.58 & $-53$ 18 59.6 & $4.1 \pm 0.4$ & $3.9^{+0.5}_{-0.5}$ & 10.0 \\
AzTEC J044658.41$-$531534.1& ADFS-AzTEC21  & 4 46 58.41 & $-53$ 15 34.1 & $5.0 \pm 0.5$ & $4.7^{+0.6}_{-0.5}$ & 9.9  \\
AzTEC J044455.57$-$533423.0& ADFS-AzTEC22  & 4 44 55.57 & $-53$ 34 23.0 & $3.9 \pm 0.4$ & $3.7^{+0.5}_{-0.4}$ & 9.8  \\
AzTEC J044540.95$-$533318.2& ADFS-AzTEC23  & 4 45 40.95 & $-53$ 33 18.2 & $3.2 \pm 0.3$ & $3.0^{+0.4}_{-0.4}$ & 9.7  \\
AzTEC J044522.00$-$532700.9& ADFS-AzTEC24  & 4 45 22.00 & $-53$ 27 00.9 & $3.6 \pm 0.4$ & $3.4^{+0.4}_{-0.4}$ & 9.7  \\
AzTEC J044659.94$-$531907.3& ADFS-AzTEC25  & 4 46 59.94 & $-53$ 19 07.3 & $4.0 \pm 0.4$ & $3.7^{+0.5}_{-0.5}$ & 9.7  \\
AzTEC J044629.83$-$533345.0& ADFS-AzTEC26  & 4 46 29.83 & $-53$ 33 45.0 & $3.3 \pm 0.4$ & $3.1^{+0.4}_{-0.4}$ & 9.3  \\
AzTEC J044658.48$-$532327.8& ADFS-AzTEC27  & 4 46 58.48 & $-53$ 23 27.8 & $3.6 \pm 0.4$ & $3.4^{+0.4}_{-0.5}$ & 9.1  \\
AzTEC J044612.33$-$532743.6& ADFS-AzTEC28  & 4 46 12.33 & $-53$ 27 43.6 & $2.9 \pm 0.3$ & $2.7^{+0.4}_{-0.4}$ & 9.1  \\
AzTEC J044714.90$-$531738.9& ADFS-AzTEC29  & 4 47 14.90 & $-53$ 17 38.9 & $3.7 \pm 0.4$ & $3.4^{+0.4}_{-0.5}$ & 9.1  \\
AzTEC J044805.34$-$532433.2& ADFS-AzTEC30  & 4 48 05.34 & $-53$ 24 33.2 & $4.4 \pm 0.5$ & $4.1^{+0.6}_{-0.5}$ & 9.0  \\
AzTEC J044351.90$-$534443.3& ADFS-AzTEC31  & 4 43 51.90 & $-53$ 44 43.3 & $4.8 \pm 0.5$ & $4.5^{+0.6}_{-0.6}$ & 8.8  \\
AzTEC J044553.57$-$533520.6& ADFS-AzTEC32  & 4 45 53.57 & $-53$ 35 20.6 & $2.8 \pm 0.3$ & $2.6^{+0.4}_{-0.4}$ & 8.7  \\
AzTEC J044706.39$-$531612.5& ADFS-AzTEC33  & 4 47 06.39 & $-53$ 16 12.5 & $4.0 \pm 0.5$ & $3.7^{+0.5}_{-0.5}$ & 8.7  \\
AzTEC J044538.56$-$532827.6& ADFS-AzTEC34  & 4 45 38.56 & $-53$ 28 27.6 & $2.8 \pm 0.3$ & $2.6^{+0.4}_{-0.4}$ & 8.4  \\
AzTEC J044853.11$-$531557.9& ADFS-AzTEC35  & 4 48 53.11 & $-53$ 15 57.9 & $4.1 \pm 0.5$ & $3.7^{+0.5}_{-0.6}$ & 8.2  \\
AzTEC J044556.43$-$533929.7& ADFS-AzTEC36  & 4 45 56.43 & $-53$ 39 29.7 & $3.5 \pm 0.4$ & $3.3^{+0.5}_{-0.5}$ & 8.1  \\
AzTEC J044348.44$-$534314.2& ADFS-AzTEC37  & 4 43 48.44 & $-53$ 43 14.2 & $4.3 \pm 0.5$ & $3.9^{+0.6}_{-0.6}$ & 8.0  \\
AzTEC J044824.87$-$531501.3& ADFS-AzTEC38  & 4 48 24.87 & $-53$ 15 01.3 & $3.5 \pm 0.4$ & $3.2^{+0.5}_{-0.5}$ & 8.0  \\
AzTEC J044609.02$-$532710.8& ADFS-AzTEC39  & 4 46 09.02 & $-53$ 27 10.8 & $2.6 \pm 0.3$ & $2.4^{+0.4}_{-0.4}$ & 8.0  \\
AzTEC J044534.14$-$533939.6& ADFS-AzTEC40  & 4 45 34.14 & $-53$ 39 39.6 & $3.4 \pm 0.4$ & $3.1^{+0.5}_{-0.5}$ & 7.9  \\
AzTEC J044354.85$-$533928.8& ADFS-AzTEC41  & 4 43 54.85 & $-53$ 39 28.8 & $3.9 \pm 0.5$ & $3.6^{+0.6}_{-0.5}$ & 7.9  \\
AzTEC J044630.39$-$533154.7& ADFS-AzTEC42  & 4 46 30.39 & $-53$ 31 54.7 & $2.7 \pm 0.3$ & $2.5^{+0.4}_{-0.4}$ & 7.8  \\
AzTEC J044607.77$-$532813.4& ADFS-AzTEC43  & 4 46 07.77 & $-53$ 28 13.4 & $2.5 \pm 0.3$ & $2.3^{+0.4}_{-0.3}$ & 7.8  \\
AzTEC J044643.79$-$533002.4& ADFS-AzTEC44  & 4 46 43.79 & $-53$ 30 02.4 & $3.0 \pm 0.4$ & $2.7^{+0.4}_{-0.5}$ & 7.6  \\
AzTEC J044451.57$-$533544.1& ADFS-AzTEC45  & 4 44 51.57 & $-53$ 35 44.1 & $3.1 \pm 0.4$ & $2.9^{+0.5}_{-0.5}$ & 7.5  \\
AzTEC J044420.88$-$534011.4& ADFS-AzTEC46  & 4 44 20.88 & $-53$ 40 11.4 & $3.4 \pm 0.5$ & $3.1^{+0.5}_{-0.5}$ & 7.3  \\
AzTEC J044541.74$-$533445.2& ADFS-AzTEC47  & 4 45 41.74 & $-53$ 34 45.2 & $2.3 \pm 0.3$ & $2.1^{+0.4}_{-0.4}$ & 7.3  \\
AzTEC J044437.74$-$534637.8& ADFS-AzTEC48  & 4 44 37.74 & $-53$ 46 37.8 & $3.8 \pm 0.5$ & $3.4^{+0.6}_{-0.6}$ & 7.2  \\
AzTEC J044528.39$-$533715.8& ADFS-AzTEC49  & 4 45 28.39 & $-53$ 37 15.8 & $2.4 \pm 0.3$ & $2.2^{+0.4}_{-0.4}$ & 7.1  \\
AzTEC J044456.63$-$533616.2& ADFS-AzTEC50  & 4 44 56.63 & $-53$ 36 16.2 & $2.8 \pm 0.4$ & $2.5^{+0.4}_{-0.5}$ & 7.1  \\
AzTEC J044623.49$-$533552.4& ADFS-AzTEC51  & 4 46 23.49 & $-53$ 35 52.4 & $2.6 \pm 0.4$ & $2.4^{+0.4}_{-0.4}$ & 6.9  \\
AzTEC J044344.41$-$534325.9& ADFS-AzTEC52  & 4 43 44.41 & $-53$ 43 25.9 & $3.8 \pm 0.6$ & $3.4^{+0.6}_{-0.6}$ & 6.9  \\
AzTEC J044742.49$-$531951.2& ADFS-AzTEC53  & 4 47 42.49 & $-53$ 19 51.2 & $2.4 \pm 0.4$ & $2.2^{+0.4}_{-0.4}$ & 6.9  \\
AzTEC J044721.15$-$532659.2& ADFS-AzTEC54  & 4 47 21.15 & $-53$ 26 59.2 & $2.9 \pm 0.4$ & $2.6^{+0.5}_{-0.5}$ & 6.8  \\
AzTEC J044423.80$-$533413.5& ADFS-AzTEC55  & 4 44 23.80 & $-53$ 34 13.5 & $3.7 \pm 0.6$ & $3.3^{+0.6}_{-0.6}$ & 6.8  \\
\hline
\end{tabular}
\end{table*}

\begin{table*}
\contcaption{}
\centering
\begin{tabular}{@{}clccccc@{}}
\hline
Name & 
\multicolumn{1}{c}{ID} & 
R.A. & 
Dec. & 
$S_{\rm observed}$ &
$S_{\rm{deboosted}}$ &
S/N \\
& & ($^h$ $^m$ $^s$) & ($^{\circ}$ $'$ $''$) & (mJy) & (mJy) & \\
\hline
AzTEC J044435.35$-$534346.6& ADFS-AzTEC56  & 4 44 35.35 & $-53$ 43 46.6 & $3.1 \pm 0.5$ & $2.8^{+0.5}_{-0.5}$ & 6.7  \\
AzTEC J044611.47$-$532152.8& ADFS-AzTEC57  & 4 46 11.47 & $-53$ 21 52.8 & $3.5 \pm 0.5$ & $3.1^{+0.5}_{-0.6}$ & 6.7  \\
AzTEC J044516.36$-$533309.9& ADFS-AzTEC58  & 4 45 16.36 & $-53$ 33 09.9 & $2.2 \pm 0.3$ & $2.0^{+0.4}_{-0.4}$ & 6.7  \\
AzTEC J044639.12$-$532518.6& ADFS-AzTEC59  & 4 46 39.12 & $-53$ 25 18.6 & $2.5 \pm 0.4$ & $2.2^{+0.4}_{-0.4}$ & 6.7  \\
AzTEC J044527.26$-$533518.6& ADFS-AzTEC60  & 4 45 27.26 & $-53$ 35 18.6 & $2.1 \pm 0.3$ & $1.9^{+0.4}_{-0.4}$ & 6.7  \\
AzTEC J044858.01$-$531524.3& ADFS-AzTEC61  & 4 48 58.01 & $-53$ 15 24.3 & $3.5 \pm 0.5$ & $3.2^{+0.6}_{-0.6}$ & 6.6  \\
AzTEC J044722.55$-$531825.8& ADFS-AzTEC62  & 4 47 22.55 & $-53$ 18 25.8 & $2.5 \pm 0.4$ & $2.3^{+0.5}_{-0.4}$ & 6.6  \\
AzTEC J044400.63$-$534228.2& ADFS-AzTEC63  & 4 44 00.63 & $-53$ 42 28.2 & $3.3 \pm 0.5$ & $2.9^{+0.5}_{-0.6}$ & 6.6  \\
AzTEC J044813.94$-$532244.7& ADFS-AzTEC64  & 4 48 13.94 & $-53$ 22 44.7 & $3.0 \pm 0.5$ & $2.7^{+0.5}_{-0.5}$ & 6.6  \\
AzTEC J044822.25$-$531420.0& ADFS-AzTEC65  & 4 48 22.25 & $-53$ 14 20.0 & $2.9 \pm 0.4$ & $2.6^{+0.5}_{-0.5}$ & 6.5  \\
AzTEC J044503.53$-$532816.7& ADFS-AzTEC66  & 4 45 03.53 & $-53$ 28 16.7 & $2.8 \pm 0.4$ & $2.5^{+0.5}_{-0.5}$ & 6.5  \\
AzTEC J044826.06$-$531413.7& ADFS-AzTEC67  & 4 48 26.06 & $-53$ 14 13.7 & $2.9 \pm 0.4$ & $2.6^{+0.5}_{-0.5}$ & 6.5  \\
AzTEC J044857.68$-$531827.0& ADFS-AzTEC68  & 4 48 57.68 & $-53$ 18 27.0 & $3.5 \pm 0.5$ & $3.1^{+0.6}_{-0.6}$ & 6.5  \\
AzTEC J044556.20$-$532541.2& ADFS-AzTEC69  & 4 45 56.20 & $-53$ 25 41.2 & $2.2 \pm 0.4$ & $2.0^{+0.4}_{-0.4}$ & 6.2  \\
AzTEC J044603.17$-$533847.5& ADFS-AzTEC70  & 4 46 03.17 & $-53$ 38 47.5 & $2.6 \pm 0.4$ & $2.3^{+0.5}_{-0.4}$ & 6.2  \\
AzTEC J044358.65$-$533731.2& ADFS-AzTEC71  & 4 43 58.65 & $-53$ 37 31.2 & $3.1 \pm 0.5$ & $2.7^{+0.5}_{-0.6}$ & 6.2  \\
AzTEC J044717.48$-$532614.8& ADFS-AzTEC72  & 4 47 17.48 & $-53$ 26 14.8 & $2.5 \pm 0.4$ & $2.2^{+0.5}_{-0.4}$ & 6.1  \\
AzTEC J044428.74$-$534129.4& ADFS-AzTEC73  & 4 44 28.74 & $-53$ 41 29.4 & $2.8 \pm 0.5$ & $2.4^{+0.5}_{-0.5}$ & 6.0  \\
AzTEC J044747.96$-$531305.5& ADFS-AzTEC74  & 4 47 47.96 & $-53$ 13 05.5 & $3.1 \pm 0.5$ & $2.7^{+0.6}_{-0.6}$ & 6.0  \\
AzTEC J044545.17$-$533815.3& ADFS-AzTEC75  & 4 45 45.17 & $-53$ 38 15.3 & $2.2 \pm 0.4$ & $1.9^{+0.4}_{-0.4}$ & 6.0  \\
AzTEC J044502.41$-$533934.4& ADFS-AzTEC76  & 4 45 02.41 & $-53$ 39 34.4 & $2.5 \pm 0.4$ & $2.2^{+0.5}_{-0.5}$ & 6.0  \\
AzTEC J044554.73$-$533829.6& ADFS-AzTEC77  & 4 45 54.73 & $-53$ 38 29.6 & $2.3 \pm 0.4$ & $2.0^{+0.5}_{-0.4}$ & 5.9  \\
AzTEC J044744.88$-$531622.8& ADFS-AzTEC78  & 4 47 44.88 & $-53$ 16 22.8 & $2.3 \pm 0.4$ & $2.0^{+0.4}_{-0.5}$ & 5.9  \\
AzTEC J044614.68$-$532631.6& ADFS-AzTEC79  & 4 46 14.68 & $-53$ 26 31.6 & $2.0 \pm 0.3$ & $1.7^{+0.4}_{-0.4}$ & 5.9  \\
AzTEC J044635.07$-$532950.7& ADFS-AzTEC80  & 4 46 35.07 & $-53$ 29 50.7 & $2.1 \pm 0.4$ & $1.8^{+0.4}_{-0.4}$ & 5.8  \\
AzTEC J044409.74$-$534607.4& ADFS-AzTEC81  & 4 44 09.74 & $-53$ 46 07.4 & $3.1 \pm 0.5$ & $2.7^{+0.6}_{-0.6}$ & 5.8  \\
AzTEC J044432.70$-$534419.3& ADFS-AzTEC82  & 4 44 32.70 & $-53$ 44 19.3 & $2.7 \pm 0.5$ & $2.4^{+0.6}_{-0.5}$ & 5.8  \\
AzTEC J044429.65$-$534659.4& ADFS-AzTEC83  & 4 44 29.65 & $-53$ 46 59.4 & $3.1 \pm 0.5$ & $2.7^{+0.6}_{-0.6}$ & 5.8  \\
AzTEC J044531.48$-$533019.1& ADFS-AzTEC84  & 4 45 31.48 & $-53$ 30 19.1 & $1.9 \pm 0.3$ & $1.6^{+0.4}_{-0.4}$ & 5.7  \\
AzTEC J044553.94$-$532908.4& ADFS-AzTEC85  & 4 45 53.94 & $-53$ 29 08.4 & $1.8 \pm 0.3$ & $1.6^{+0.4}_{-0.4}$ & 5.6  \\
AzTEC J044434.71$-$534125.5& ADFS-AzTEC86  & 4 44 34.71 & $-53$ 41 25.5 & $2.5 \pm 0.4$ & $2.2^{+0.5}_{-0.5}$ & 5.6  \\
AzTEC J044738.06$-$532754.8& ADFS-AzTEC87  & 4 47 38.06 & $-53$ 27 54.8 & $2.9 \pm 0.5$ & $2.5^{+0.6}_{-0.6}$ & 5.5  \\
AzTEC J044439.10$-$533704.0& ADFS-AzTEC88  & 4 44 39.10 & $-53$ 37 04.0 & $2.5 \pm 0.5$ & $2.2^{+0.5}_{-0.5}$ & 5.5  \\
AzTEC J044642.77$-$532926.9& ADFS-AzTEC89  & 4 46 42.77 & $-53$ 29 26.9 & $2.1 \pm 0.4$ & $1.8^{+0.4}_{-0.5}$ & 5.5  \\
AzTEC J044732.69$-$531422.0& ADFS-AzTEC90  & 4 47 32.69 & $-53$ 14 22.0 & $2.6 \pm 0.5$ & $2.2^{+0.5}_{-0.5}$ & 5.4  \\
AzTEC J044509.91$-$532831.2& ADFS-AzTEC91  & 4 45 09.91 & $-53$ 28 31.2 & $2.1 \pm 0.4$ & $1.8^{+0.4}_{-0.5}$ & 5.4  \\
AzTEC J044749.46$-$532244.4& ADFS-AzTEC92  & 4 47 49.46 & $-53$ 22 44.4 & $2.0 \pm 0.4$ & $1.8^{+0.5}_{-0.4}$ & 5.4  \\
AzTEC J044438.94$-$533316.8& ADFS-AzTEC93  & 4 44 38.94 & $-53$ 33 16.8 & $2.9 \pm 0.5$ & $2.4^{+0.6}_{-0.6}$ & 5.4  \\
AzTEC J044504.27$-$533030.9& ADFS-AzTEC94  & 4 45 04.27 & $-53$ 30 30.9 & $2.0 \pm 0.4$ & $1.8^{+0.4}_{-0.4}$ & 5.3  \\
AzTEC J044632.65$-$533457.6& ADFS-AzTEC95  & 4 46 32.65 & $-53$ 34 57.6 & $2.1 \pm 0.4$ & $1.8^{+0.4}_{-0.5}$ & 5.3  \\
AzTEC J044751.90$-$531920.3& ADFS-AzTEC96  & 4 47 51.90 & $-53$ 19 20.3 & $1.9 \pm 0.4$ & $1.6^{+0.4}_{-0.4}$ & 5.3  \\
AzTEC J044627.13$-$533224.4& ADFS-AzTEC97  & 4 46 27.13 & $-53$ 32 24.4 & $1.8 \pm 0.3$ & $1.5^{+0.4}_{-0.4}$ & 5.2  \\
AzTEC J044607.98$-$533416.2& ADFS-AzTEC98  & 4 46 07.98 & $-53$ 34 16.2 & $1.7 \pm 0.3$ & $1.5^{+0.4}_{-0.4}$ & 5.2  \\
AzTEC J044537.32$-$532311.4& ADFS-AzTEC99  & 4 45 37.32 & $-53$ 23 11.4 & $2.8 \pm 0.5$ & $2.3^{+0.6}_{-0.6}$ & 5.2  \\
AzTEC J044608.37$-$533550.4& ADFS-AzTEC100 & 4 46 08.37 & $-53$ 35 50.4 & $1.7 \pm 0.3$ & $1.5^{+0.4}_{-0.4}$ & 5.1  \\
AzTEC J044654.55$-$532308.3& ADFS-AzTEC101 & 4 46 54.55 & $-53$ 23 08.3 & $2.0 \pm 0.4$ & $1.8^{+0.5}_{-0.5}$ & 5.1  \\
AzTEC J044747.12$-$532011.2& ADFS-AzTEC102 & 4 47 47.12 & $-53$ 20 11.2 & $1.8 \pm 0.4$ & $1.6^{+0.4}_{-0.4}$ & 5.1  \\
AzTEC J044526.81$-$533031.2& ADFS-AzTEC103 & 4 45 26.81 & $-53$ 30 31.2 & $1.7 \pm 0.3$ & $1.4^{+0.4}_{-0.4}$ & 5.1  \\
AzTEC J044628.14$-$533254.4& ADFS-AzTEC104 & 4 46 28.14 & $-53$ 32 54.4 & $1.7 \pm 0.3$ & $1.5^{+0.4}_{-0.4}$ & 5.1  \\
AzTEC J044837.01$-$531924.2& ADFS-AzTEC105 & 4 48 37.01 & $-53$ 19 24.2 & $2.3 \pm 0.5$ & $2.0^{+0.5}_{-0.5}$ & 5.0  \\
AzTEC J044558.64$-$533747.4& ADFS-AzTEC106 & 4 45 58.64 & $-53$ 37 47.4 & $1.8 \pm 0.4$ & $1.6^{+0.4}_{-0.4}$ & 5.0  \\
AzTEC J044821.59$-$531219.6& ADFS-AzTEC107 & 4 48 21.59 & $-53$ 12 19.6 & $2.5 \pm 0.5$ & $2.1^{+0.6}_{-0.5}$ & 5.0  \\
AzTEC J044741.65$-$531839.1& ADFS-AzTEC108 & 4 47 41.65 & $-53$ 18 39.1 & $1.8 \pm 0.4$ & $1.5^{+0.4}_{-0.4}$ & 5.0  \\
AzTEC J044442.20$-$534413.9& ADFS-AzTEC109 & 4 44 42.20 & $-53$ 44 13.9 & $2.4 \pm 0.5$ & $2.0^{+0.5}_{-0.6}$ & 5.0  \\
AzTEC J044720.44$-$532526.5& ADFS-AzTEC110 & 4 47 20.44 & $-53$ 25 26.5 & $2.0 \pm 0.4$ & $1.7^{+0.4}_{-0.5}$ & 5.0  \\
AzTEC J044543.78$-$532612.1& ADFS-AzTEC111 & 4 45 43.78 & $-53$ 26 12.1 & $1.7 \pm 0.3$ & $1.5^{+0.4}_{-0.4}$ & 4.9  \\
AzTEC J044403.61$-$533816.8& ADFS-AzTEC112 & 4 44 03.61 & $-53$ 38 16.8 & $2.4 \pm 0.5$ & $2.0^{+0.6}_{-0.5}$ & 4.9  \\
\hline
\end{tabular}
\end{table*}

\begin{table*}
\contcaption{}
\centering
\begin{tabular}{@{}clccccc@{}}
\hline
Name & 
\multicolumn{1}{c}{ID} & 
R.A. & 
Dec. & 
$S_{\rm observed}$ &
$S_{\rm{deboosted}}$ &
S/N \\
& & ($^h$ $^m$ $^s$) & ($^{\circ}$ $'$ $''$) & (mJy) & (mJy) & \\
\hline
AzTEC J044517.36$-$532619.7& ADFS-AzTEC113 & 4 45 17.36 & $-53$ 26 19.7 & $2.0 \pm 0.4$ & $1.7^{+0.5}_{-0.5}$ & 4.9 \\
AzTEC J044511.45$-$533328.2& ADFS-AzTEC114 & 4 45 11.45 & $-53$ 33 28.2 & $1.6 \pm 0.3$ & $1.4^{+0.4}_{-0.4}$ & 4.9 \\
AzTEC J044834.22$-$531727.4& ADFS-AzTEC115 & 4 48 34.22 & $-53$ 17 27.4 & $2.1 \pm 0.4$ & $1.8^{+0.5}_{-0.5}$ & 4.9 \\
AzTEC J044759.97$-$531235.0& ADFS-AzTEC116 & 4 47 59.97 & $-53$ 12 35.0 & $2.5 \pm 0.5$ & $2.1^{+0.6}_{-0.6}$ & 4.8 \\
AzTEC J044540.52$-$532927.6& ADFS-AzTEC117 & 4 45 40.52 & $-53$ 29 27.6 & $1.6 \pm 0.3$ & $1.3^{+0.4}_{-0.4}$ & 4.8 \\
AzTEC J044529.40$-$534100.9& ADFS-AzTEC118 & 4 45 29.40 & $-53$ 41 00.9 & $2.4 \pm 0.5$ & $2.0^{+0.6}_{-0.6}$ & 4.7 \\
AzTEC J044553.28$-$533017.6& ADFS-AzTEC119 & 4 45 53.28 & $-53$ 30 17.6 & $1.5 \pm 0.3$ & $1.3^{+0.4}_{-0.4}$ & 4.7 \\
AzTEC J044634.10$-$533721.6& ADFS-AzTEC120 & 4 46 34.10 & $-53$ 37 21.6 & $2.4 \pm 0.5$ & $2.0^{+0.6}_{-0.6}$ & 4.7 \\
AzTEC J044515.32$-$533354.3& ADFS-AzTEC121 & 4 45 15.32 & $-53$ 33 54.3 & $1.5 \pm 0.3$ & $1.3^{+0.4}_{-0.4}$ & 4.7 \\
AzTEC J044435.41$-$533529.4& ADFS-AzTEC122 & 4 44 35.41 & $-53$ 35 29.4 & $2.2 \pm 0.5$ & $1.8^{+0.5}_{-0.6}$ & 4.6 \\
AzTEC J044458.94$-$533504.3& ADFS-AzTEC123 & 4 44 58.94 & $-53$ 35 04.3 & $1.7 \pm 0.4$ & $1.4^{+0.4}_{-0.5}$ & 4.5 \\
AzTEC J044636.36$-$532915.4& ADFS-AzTEC124 & 4 46 36.36 & $-53$ 29 15.4 & $1.6 \pm 0.4$ & $1.3^{+0.4}_{-0.4}$ & 4.5 \\
AzTEC J044605.18$-$533137.8& ADFS-AzTEC125 & 4 46 05.18 & $-53$ 31 37.8 & $1.5 \pm 0.3$ & $1.2^{+0.4}_{-0.4}$ & 4.5 \\
AzTEC J044640.93$-$532608.2& ADFS-AzTEC126 & 4 46 40.93 & $-53$ 26 08.2 & $1.6 \pm 0.4$ & $1.3^{+0.4}_{-0.4}$ & 4.4 \\
AzTEC J044634.09$-$532612.2& ADFS-AzTEC127 & 4 46 34.09 & $-53$ 26 12.2 & $1.6 \pm 0.4$ & $1.3^{+0.4}_{-0.5}$ & 4.4 \\
AzTEC J044756.14$-$531727.7& ADFS-AzTEC128 & 4 47 56.14 & $-53$ 17 27.7 & $1.6 \pm 0.4$ & $1.4^{+0.5}_{-0.4}$ & 4.3 \\
AzTEC J044609.04$-$532913.4& ADFS-AzTEC129 & 4 46 09.04 & $-53$ 29 13.4 & $1.4 \pm 0.3$ & $1.1^{+0.4}_{-0.4}$ & 4.3 \\
AzTEC J044756.92$-$531637.4& ADFS-AzTEC130 & 4 47 56.92 & $-53$ 16 37.4 & $1.7 \pm 0.4$ & $1.4^{+0.5}_{-0.4}$ & 4.3 \\
AzTEC J044528.27$-$533551.3& ADFS-AzTEC131 & 4 45 28.27 & $-53$ 35 51.3 & $1.4 \pm 0.3$ & $1.1^{+0.4}_{-0.4}$ & 4.3 \\
AzTEC J044532.76$-$532921.6& ADFS-AzTEC132 & 4 45 32.76 & $-53$ 29 21.6 & $1.4 \pm 0.3$ & $1.2^{+0.4}_{-0.4}$ & 4.3 \\
AzTEC J044551.09$-$532505.4& ADFS-AzTEC133 & 4 45 51.09 & $-53$ 25 05.4 & $1.6 \pm 0.4$ & $1.3^{+0.4}_{-0.5}$ & 4.3 \\
AzTEC J044438.82$-$534047.6& ADFS-AzTEC134 & 4 44 38.82 & $-53$ 40 47.6 & $1.9 \pm 0.4$ & $1.5^{+0.5}_{-0.5}$ & 4.3 \\
AzTEC J044449.52$-$533723.0& ADFS-AzTEC135 & 4 44 49.52 & $-53$ 37 23.0 & $1.8 \pm 0.4$ & $1.5^{+0.5}_{-0.5}$ & 4.3 \\
AzTEC J044723.01$-$531414.1& ADFS-AzTEC136 & 4 47 23.01 & $-53$ 14 14.1 & $2.2 \pm 0.5$ & $1.7^{+0.6}_{-0.6}$ & 4.2 \\
AzTEC J044746.74$-$532129.6& ADFS-AzTEC137 & 4 47 46.74 & $-53$ 21 29.6 & $1.5 \pm 0.4$ & $1.3^{+0.4}_{-0.4}$ & 4.2 \\
AzTEC J044522.17$-$533345.0& ADFS-AzTEC138 & 4 45 22.17 & $-53$ 33 45.0 & $1.3 \pm 0.3$ & $1.1^{+0.4}_{-0.4}$ & 4.2 \\
AzTEC J044351.18$-$534228.7& ADFS-AzTEC139 & 4 43 51.18 & $-53$ 42 28.7 & $2.2 \pm 0.5$ & $1.7^{+0.6}_{-0.6}$ & 4.2 \\
AzTEC J044758.09$-$531825.1& ADFS-AzTEC140 & 4 47 58.09 & $-53$ 18 25.1 & $1.6 \pm 0.4$ & $1.3^{+0.5}_{-0.4}$ & 4.2 \\
AzTEC J044733.34$-$532222.1& ADFS-AzTEC141 & 4 47 33.34 & $-53$ 22 22.1 & $1.6 \pm 0.4$ & $1.3^{+0.5}_{-0.4}$ & 4.2 \\
AzTEC J044746.06$-$532623.4& ADFS-AzTEC142 & 4 47 46.06 & $-53$ 26 23.4 & $2.0 \pm 0.5$ & $1.6^{+0.5}_{-0.6}$ & 4.1 \\
AzTEC J044715.12$-$532500.9& ADFS-AzTEC143 & 4 47 15.12 & $-53$ 25 00.9 & $1.6 \pm 0.4$ & $1.3^{+0.5}_{-0.5}$ & 4.1 \\
AzTEC J044458.75$-$534319.5& ADFS-AzTEC144 & 4 44 58.75 & $-53$ 43 19.5 & $2.1 \pm 0.5$ & $1.6^{+0.6}_{-0.6}$ & 4.1 \\
AzTEC J044625.74$-$531931.6& ADFS-AzTEC145 & 4 46 25.74 & $-53$ 19 31.6 & $2.1 \pm 0.5$ & $1.7^{+0.6}_{-0.6}$ & 4.1 \\
AzTEC J044637.83$-$531851.1& ADFS-AzTEC146 & 4 46 37.83 & $-53$ 18 51.1 & $1.8 \pm 0.5$ & $1.4^{+0.5}_{-0.6}$ & 4.0 \\
AzTEC J044524.83$-$533134.0& ADFS-AzTEC147 & 4 45 24.83 & $-53$ 31 34.0 & $1.3 \pm 0.3$ & $1.0^{+0.4}_{-0.4}$ & 3.9 \\
AzTEC J044608.77$-$533716.8& ADFS-AzTEC148 & 4 46 08.77 & $-53$ 37 16.8 & $1.4 \pm 0.4$ & $1.1^{+0.4}_{-0.4}$ & 3.9 \\
AzTEC J044515.33$-$532955.2& ADFS-AzTEC149 & 4 45 15.33 & $-53$ 29 55.2 & $1.3 \pm 0.3$ & $1.1^{+0.4}_{-0.4}$ & 3.9 \\
AzTEC J044654.28$-$533116.4& ADFS-AzTEC150 & 4 46 54.28 & $-53$ 31 16.4 & $1.9 \pm 0.5$ & $1.5^{+0.6}_{-0.6}$ & 3.9 \\
AzTEC J044634.68$-$532251.6& ADFS-AzTEC151 & 4 46 34.68 & $-53$ 22 51.6 & $1.5 \pm 0.4$ & $1.2^{+0.5}_{-0.5}$ & 3.8 \\
AzTEC J044508.10$-$534204.2& ADFS-AzTEC152 & 4 45 08.10 & $-53$ 42 04.2 & $2.0 \pm 0.5$ & $1.5^{+0.6}_{-0.6}$ & 3.8 \\
AzTEC J044456.83$-$533216.7& ADFS-AzTEC153 & 4 44 56.83 & $-53$ 32 16.7 & $1.5 \pm 0.4$ & $1.2^{+0.5}_{-0.5}$ & 3.8 \\
AzTEC J044629.37$-$532922.3& ADFS-AzTEC154 & 4 46 29.37 & $-53$ 29 22.3 & $1.3 \pm 0.3$ & $1.0^{+0.4}_{-0.4}$ & 3.8 \\
AzTEC J044655.70$-$532707.5& ADFS-AzTEC155 & 4 46 55.70 & $-53$ 27 07.5 & $1.5 \pm 0.4$ & $1.2^{+0.5}_{-0.5}$ & 3.8 \\
AzTEC J044834.45$-$531642.0& ADFS-AzTEC156 & 4 48 34.45 & $-53$ 16 42.0 & $1.6 \pm 0.4$ & $1.3^{+0.5}_{-0.5}$ & 3.8 \\
AzTEC J044346.80$-$534131.8& ADFS-AzTEC157 & 4 43 46.80 & $-53$ 41 31.8 & $2.0 \pm 0.5$ & $1.5^{+0.7}_{-0.6}$ & 3.8 \\
AzTEC J044822.88$-$532151.5& ADFS-AzTEC158 & 4 48 22.88 & $-53$ 21 51.5 & $1.7 \pm 0.5$ & $1.3^{+0.5}_{-0.6}$ & 3.8 \\
AzTEC J044533.34$-$532423.6& ADFS-AzTEC159 & 4 45 33.34 & $-53$ 24 23.6 & $1.7 \pm 0.4$ & $1.3^{+0.6}_{-0.5}$ & 3.7 \\
AzTEC J044522.66$-$532518.8& ADFS-AzTEC160 & 4 45 22.66 & $-53$ 25 18.8 & $1.6 \pm 0.4$ & $1.2^{+0.5}_{-0.5}$ & 3.7 \\
AzTEC J044407.06$-$533950.0& ADFS-AzTEC161 & 4 44 07.06 & $-53$ 39 50.0 & $1.8 \pm 0.5$ & $1.3^{+0.6}_{-0.6}$ & 3.7 \\
AzTEC J044646.61$-$531632.6& ADFS-AzTEC162 & 4 46 46.61 & $-53$ 16 32.6 & $1.8 \pm 0.5$ & $1.3^{+0.6}_{-0.6}$ & 3.7 \\
AzTEC J044516.25$-$534025.3& ADFS-AzTEC163 & 4 45 16.25 & $-53$ 40 25.3 & $1.7 \pm 0.5$ & $1.2^{+0.5}_{-0.5}$ & 3.7 \\
AzTEC J044512.72$-$533644.8& ADFS-AzTEC164 & 4 45 12.72 & $-53$ 36 44.8 & $1.3 \pm 0.4$ & $1.0^{+0.4}_{-0.5}$ & 3.7 \\
AzTEC J044738.40$-$532327.9& ADFS-AzTEC165 & 4 47 38.40 & $-53$ 23 27.9 & $1.4 \pm 0.4$ & $1.1^{+0.5}_{-0.5}$ & 3.6 \\
AzTEC J044719.99$-$532902.4& ADFS-AzTEC166 & 4 47 19.99 & $-53$ 29 02.4 & $1.8 \pm 0.5$ & $1.3^{+0.6}_{-0.6}$ & 3.6 \\
AzTEC J044422.24$-$534544.5& ADFS-AzTEC167 & 4 44 22.24 & $-53$ 45 44.5 & $1.8 \pm 0.5$ & $1.3^{+0.6}_{-0.6}$ & 3.6 \\
AzTEC J044410.42$-$533838.0& ADFS-AzTEC168 & 4 44 10.42 & $-53$ 38 38.0 & $1.7 \pm 0.5$ & $1.2^{+0.5}_{-0.6}$ & 3.5 \\
AzTEC J044701.87$-$532915.7& ADFS-AzTEC169 & 4 47 01.87 & $-53$ 29 15.7 & $1.7 \pm 0.5$ & $1.2^{+0.5}_{-0.6}$ & 3.5 \\
\hline
\end{tabular}
\end{table*}
\begin{table*}
\contcaption{}
\centering
\begin{tabular}{@{}clccccc@{}}
\hline
Name & 
\multicolumn{1}{c}{ID} & 
R.A. & 
Dec. & 
$S_{\rm observed}$ &
$S_{\rm{deboosted}}$ &
S/N \\
& & ($^h$ $^m$ $^s$) & ($^{\circ}$ $'$ $''$) & (mJy) & (mJy) & \\
\hline
AzTEC J044327.57$-$534158.07& ADFS-AzTEC170 & 4 43 27.57 & $-53$ 41 58.07 & $6.6 \pm 0.7$ & $6.1^{+0.8}_{-0.7}$ & 9.5 \\
AzTEC J044435.40$-$533316.47& ADFS-AzTEC171 & 4 44 35.40 & $-53$ 33 16.47 & $4.4 \pm 0.6$ & $4.0^{+0.6}_{-0.6}$ & 7.8 \\
AzTEC J044803.16$-$532615.99& ADFS-AzTEC172 & 4 48 03.16 & $-53$ 26 15.99 & $4.6 \pm 0.6$ & $4.2^{+0.7}_{-0.6}$ & 7.7 \\
AzTEC J044819.80$-$532444.49& ADFS-AzTEC173 & 4 48 19.80 & $-53$ 24 44.49 & $4.4 \pm 0.6$ & $3.9^{+0.6}_{-0.7}$ & 7.3 \\
AzTEC J044421.55$-$534823.46& ADFS-AzTEC174 & 4 44 21.55 & $-53$ 48 23.46 & $4.4 \pm 0.6$ & $3.9^{+0.7}_{-0.7}$ & 7.2 \\
AzTEC J044502.44$-$534452.07& ADFS-AzTEC175 & 4 45 02.44 & $-53$ 44 52.07 & $4.2 \pm 0.6$ & $3.8^{+0.7}_{-0.7}$ & 7.0 \\
AzTEC J044511.21$-$532400.67& ADFS-AzTEC176 & 4 45 11.21 & $-53$ 24 00.67 & $4.0 \pm 0.7$ & $3.4^{+0.7}_{-0.7}$ & 6.0 \\
AzTEC J044636.65$-$531618.03& ADFS-AzTEC177 & 4 46 36.65 & $-53$ 16 18.03 & $3.3 \pm 0.6$ & $2.9^{+0.7}_{-0.6}$ & 5.7 \\
AzTEC J044652.58$-$531505.07& ADFS-AzTEC178 & 4 46 52.58 & $-53$ 15 05.07 & $3.1 \pm 0.6$ & $2.6^{+0.6}_{-0.6}$ & 5.5 \\
AzTEC J044352.49$-$534647.20& ADFS-AzTEC179 & 4 43 52.49 & $-53$ 46 47.20 & $3.4 \pm 0.6$ & $2.9^{+0.7}_{-0.7}$ & 5.4 \\
AzTEC J044335.26$-$534340.94& ADFS-AzTEC180 & 4 43 35.26 & $-53$ 43 40.94 & $3.4 \pm 0.6$ & $2.8^{+0.7}_{-0.7}$ & 5.4 \\
AzTEC J044331.59$-$533937.66& ADFS-AzTEC181 & 4 43 31.59 & $-53$ 39 37.66 & $3.6 \pm 0.7$ & $3.0^{+0.8}_{-0.7}$ & 5.3 \\
AzTEC J044446.10$-$532828.82& ADFS-AzTEC182 & 4 44 46.10 & $-53$ 28 28.82 & $3.6 \pm 0.7$ & $2.9^{+0.8}_{-0.8}$ & 5.1 \\
AzTEC J044813.09$-$531023.30& ADFS-AzTEC183 & 4 48 13.09 & $-53$ 10 23.30 & $3.1 \pm 0.6$ & $2.5^{+0.7}_{-0.7}$ & 4.9 \\
AzTEC J044559.84$-$534117.11& ADFS-AzTEC184 & 4 45 59.84 & $-53$ 41 17.11 & $3.0 \pm 0.6$ & $2.4^{+0.7}_{-0.7}$ & 4.7 \\
AzTEC J044657.28$-$533204.37& ADFS-AzTEC185 & 4 46 57.28 & $-53$ 32 04.37 & $2.6 \pm 0.6$ & $2.1^{+0.7}_{-0.6}$ & 4.5 \\
AzTEC J044853.36$-$531151.36& ADFS-AzTEC186 & 4 48 53.36 & $-53$ 11 51.36 & $2.8 \pm 0.6$ & $2.2^{+0.7}_{-0.7}$ & 4.5 \\
AzTEC J044845.15$-$531020.30& ADFS-AzTEC187 & 4 48 45.15 & $-53$ 10 20.30 & $2.9 \pm 0.7$ & $2.3^{+0.8}_{-0.7}$ & 4.4 \\
AzTEC J044911.57$-$531409.24& ADFS-AzTEC188 & 4 49 11.57 & $-53$ 14 09.24 & $3.0 \pm 0.7$ & $2.3^{+0.8}_{-0.7}$ & 4.4 \\
AzTEC J044739.25$-$531234.57& ADFS-AzTEC189 & 4 47 39.25 & $-53$ 12 34.57 & $2.5 \pm 0.6$ & $1.9^{+0.7}_{-0.7}$ & 4.1 \\
AzTEC J044410.78$-$534755.51& ADFS-AzTEC190 & 4 44 10.78 & $-53$ 47 55.51 & $2.4 \pm 0.6$ & $1.8^{+0.7}_{-0.7}$ & 4.0 \\
AzTEC J044722.60$-$533044.49& ADFS-AzTEC191 & 4 47 22.60 & $-53$ 30 44.49 & $2.7 \pm 0.7$ & $2.0^{+0.8}_{-0.8}$ & 3.9 \\
AzTEC J044639.24$-$531515.80& ADFS-AzTEC192 & 4 46 39.24 & $-53$ 15 15.80 & $2.4 \pm 0.7$ & $1.7^{+0.8}_{-0.8}$ & 3.8 \\
AzTEC J044822.21$-$532519.16& ADFS-AzTEC193 & 4 48 22.21 & $-53$ 25 19.16 & $2.6 \pm 0.7$ & $1.8^{+0.8}_{-0.9}$ & 3.7 \\
AzTEC J044638.14$-$533806.11& ADFS-AzTEC194 & 4 46 38.14 & $-53$ 38 06.11 & $2.4 \pm 0.6$ & $1.7^{+0.7}_{-0.8}$ & 3.7 \\
AzTEC J044607.81$-$532028.67& ADFS-AzTEC195 & 4 46 07.81 & $-53$ 20 28.67 & $2.6 \pm 0.7$ & $1.8^{+0.8}_{-0.9}$ & 3.7 \\
AzTEC J044553.73$-$532144.72& ADFS-AzTEC196 & 4 45 53.73 & $-53$ 21 44.72 & $2.4 \pm 0.7$ & $1.6^{+0.8}_{-0.8}$ & 3.6 \\
AzTEC J044648.62$-$531447.09& ADFS-AzTEC197 & 4 46 48.62 & $-53$ 14 47.09 & $2.2 \pm 0.6$ & $1.5^{+0.8}_{-0.7}$ & 3.6 \\
AzTEC J044831.02$-$531003.43& ADFS-AzTEC198 & 4 48 31.02 & $-53$ 10 03.43 & $2.3 \pm 0.6$ & $1.5^{+0.8}_{-0.8}$ & 3.5 \\
\hline
\end{tabular}
\end{table*}

\subsection{False Detections}\label{sec:fdr}

\begin{figure}
\begin{center}
\includegraphics[width=\linewidth]{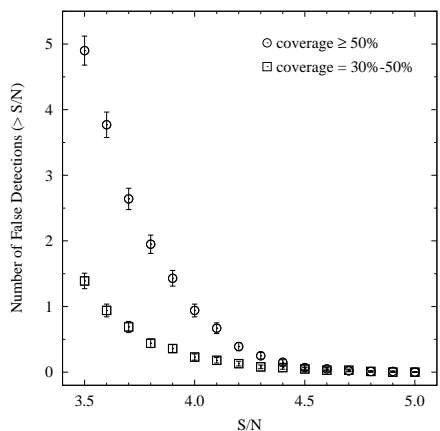}
\end{center}
\caption{
Number of false detections expected above given signal-to-noise ratio thresholds calculated in the 50\% and 30--50\% coverage regions. 
The error bars are 1$\sigma$ Poisson uncertainties. 
}
\label{fig:fdr}
\end{figure}

Monte Carlo simulations are carried out to estimate the number of spurious sources due to positive noise fluctuations. 
We conduct the standard source extraction on the 100 synthesised noise realizations, and count the number of `sources' above given S/N thresholds in steps of 0.5$\sigma$. 
Fig. \ref{fig:fdr} shows the average number of false detections as a function of S/N. 
The expected number of false detections in our $\ge$3.5$\sigma$ source catalog is $\sim$4--5 and $\sim$1--2 in the 50\% and 30\%--50\% coverage region, respectively.

\subsection{Completeness}\label{sec:completeness}

\begin{figure}
\begin{center}
\includegraphics[width=\linewidth]{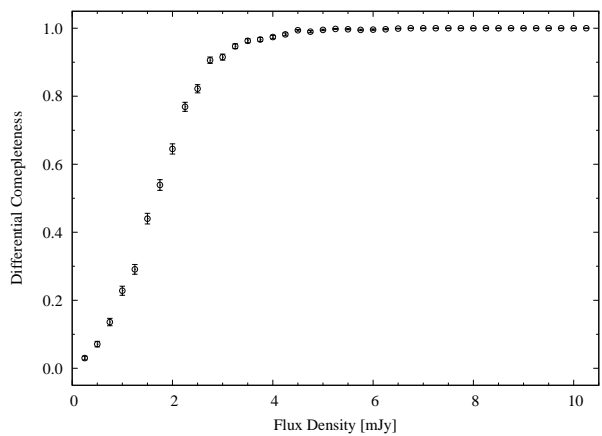}
\caption[Survey completeness]{
Survey completeness for the 50\% coverage region. 
The error bars are 1$\sigma$ from the binomial distribution. 
\label{fig:completeness}
}
\end{center}
\end{figure}

The survey completeness is computed by injecting simulated point sources of known flux densities into the real signal map one at a time. 
The input positions are randomly selected within the 50\% coverage region, but are required to be outside a $20''$ radius from a real source in the map to avoid blending. 
When a simulated source is extracted within $20''$ of its input position with S/N $\ge$ 3.5, the source is considered to be recovered. 
We repeat this 1000 times for each flux bin, and compute the fraction of output sources to input sources. 
The completeness as a function of intrinsic flux density is shown in Fig.~\ref{fig:completeness}. 
The error bars are the 68\% confidence intervals from the binomial distribution. 
The completeness is about 50\% at a flux density of 2.0~mJy.

\begin{figure}
\begin{center}
\includegraphics[width=\linewidth]{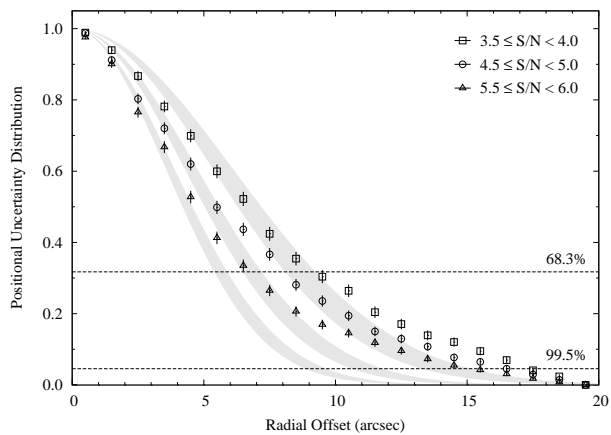}
\end{center}
\caption{
Probability that a source is detected outside an angular distance from its true position. 
The probability is calculated for sources with $3.5 \le$ SN $< 4.0$, $4.5 \le$ SN $< 5.0$, and $5.5 \le$ SN $< 6.0$. 
The horizontal dashed lines indicate 68.3\% and 99.5\% confidence intervals. 
The shaded regions represent theoretical predictions derived in \citet{ivis07} for the SN ranges. 
}
\label{fig:position}
\end{figure}

\subsection{Positional Uncertainty}\label{sec:position}

The positional error is an important indicator when identifying other-wavelength counterparts. 
The positional uncertainties for detected sources are calculated in a similar manner for the completeness calculation. 
Fake sources of known flux densities (ranging from 1 to 10~mJy in steps of 0.25~mJy) are injected into the real signal map one at a time. 
The input positions are selected randomly outside a $20''$ radius from the real sources. 
Source extractions with S/N $\ge$ 3.5 are performed within $20''$ search radius from the input positions. 
This procedures are repeated 1000 times for each flux bin, and measure the angular distances between the input and output positions. 
The probability that a source is extracted outside an angular distance from its input position is shown in Fig.~\ref{fig:position} for different S/N ranges. 
The 68.3\% confidence interval for a 3.5$\sigma$ source is $\sim$$9''$. 
We compare the positional uncertainty with theoretical predictions for uncorrelated Gaussian noise derived by \cite{ivis07}. 
Shaded regions in Fig.~\ref{fig:position} represent the theoretical probability distributions for sources with $3.5 \le$ S/N $< 4.0$, $4.5 \le$ S/N $< 5.0$, and $5.5 \le$ S/N $< 6.0$ from top to bottom. 
The positional uncertainties of AzTEC sources are broad compared to the theoretical predictions. 
It is possible that the ADF-S map is confused and the confusion noise affects source positions.

\subsection{Flux Deboosting}\label{sec:deboosting}

When dealing with a low S/N map, we need to consider the effect that flux densities of low S/N sources are boosted to above detection thresholds, due to the steep slope of number counts in the flux range of our map \citep{murd73, hogg98}. 
We correct for the flux boosting effect to estimate intrinsic flux densities of sources based on the Bayesian estimation \citep{copp05, copp06, aust09, aust10}.

To calculate the probability distribution of its intrinsic flux density (posterior flux distribution; PFD), we use the best-fit differential number counts in the ADF-S derived in \S~\ref{sec:counts_adfs}. 
We inject fake sources with flux densities ranging from $-5$ to 20 mJy into a synthesised noiseless sky at random positions. 
We iterate this process 10000 times and the prior distribution is given as the averaged flux distribution of the sources.  
The deboosted flux densities are given in Table~\ref{tab:catalog}.

\section{1.1~mm Number Counts}\label{sec:counts}

\subsection{Number Counts of the ADF-S}\label{sec:counts_adfs}

We create number counts for the 50\% coverage region, where the noise distribution is more uniform, the survey completeness at faint flux densities is high, and the number of false detections are low compared to the 30\% coverage region. 
We employ the Bayesian method, which is now commonly used for deriving number counts in millimetre/submillimetre surveys \citep[e.g.,][]{copp05, copp06, pere08, aust09, aust10, scot10}.

The PFD of each source is calculated in the same manner as used in flux deboosting (\S~\ref{sec:deboosting}). 
We adopt the best-fit Schechter function of the SCUBA/SHADES 850~$\mu$m number counts \citep{copp06} scaled to 1.1~mm as an initial prior distribution function. 
We create 20000 sample catalogs by bootstrapping off (i.e., sampling with replacement the PFDs. 
We sample only from the PFDs with a criteria of $P(S<0) \le 0.05$ as adopted in \citet{scot10}, where $P(S<0)$ is the probability that the flux densities of the detected sources deboosted to $<$0~mJy. 
The number of sources in each sample catalog is chosen randomly as a Poisson deviate from the real number of sources. 
The survey completeness is calculated by tracing output to input sources from the simulated maps. 
After correcting for the survey completeness, the mean counts and the 68.3\% confidence intervals in each flux bin with a bin size of 1~mJy are calculated from the 20000 sample catalogs. 
The deboosted fluxes in the source catalogue range from 1 to 6.4~mJy. 
Each of the derived number counts for the 20000 sample catalogs is fitted to a Schechter function of the form 
\begin{eqnarray}\label{eq:schechter}
\frac{dN}{dS} = 
N_{\rm 3 mJy}\left(\frac{S}{\rm 3\ mJy}\right)^{\alpha+1} \exp\left[\frac{-(S-{\rm 3\ mJy})}{S'}\right], 
\end{eqnarray}
where $N_{\rm 3 mJy}$ is a differential counts at 3~mJy and best-fit parameters are obtained in $S'-N_{\rm 3mJy}$ parameter space. 
We adopt the above Schechter functional form with $\alpha=-2$ since it well describes number counts derived in previous deep SMG surveys \citep[e.g.,][]{copp06, pere08, aust09, aust10, scot10}. 
The derived best-fit function is then used as a new prior distribution function and the procedure described above is repeated. 
The resultant differential and cumulative number counts are presented in Fig.~\ref{fig:dif_counts}, Fig.~\ref{fig:int_counts}, and Table \ref{tab:counts}. 
The errors indicate the 68.3\% confidence intervals. 
The best-fit parameters of the Schechter functional form in equation~(\ref{eq:schechter}) are $N_{\rm 3 mJy} = 169 \pm 19$ and $S' = 1.48 \pm 0.20$.

\begin{table}
\caption{
Differential and cumulative number counts in the ADF-S. 
The flux bin centers for the differential counts (first column) are weighted by the assumed prior. 
The errors are 68.3\% confidence intervals. 
}
\label{tab:counts}
\begin{center}
\begin{tabular}{@{}cccc@{}}
\hline\hline
Flux Density & Differential & Flux Density & Cumulative \\
(mJy) & (mJy$^{-1}$ deg$^{-2}$) & (mJy) & (deg$^{-2}$) \\
\hline
1.39 & 1074$^{+123}_{-137}$ & 1.0 & 1590$^{+134}_{-151}$ \\
2.41 &  312$^{+44 }_{-51 }$ & 2.0 &  515$^{+54 }_{-64 }$ \\
3.42 &  109$^{+24 }_{-28 }$ & 3.0 &  204$^{+31 }_{-38 }$ \\
4.42 &   63$^{+17 }_{-21 }$ & 4.0 &   95$^{+20 }_{-26 }$ \\
5.43 &   25$^{+10 }_{-14 }$ & 5.0 &   31$^{+11 }_{-15 }$ \\
6.43 &  5.7$^{+3.4}_{-5.7}$ & 6.0 &  5.8$^{+3.4}_{-5.7}$ \\
\hline
\end{tabular}
\end{center}
\end{table}

\subsection{Number Counts of the SXDF and the SSA~22 Fields Surveyed by AzTEC/ASTE}\label{sec:counts_sxdf_ssa22}

We extract number counts for two other deep fields surveyed by AzTEC on the ASTE: 
the Subaru/{\sl XMM Newton} Deep Field (SXDF) and the SSA~22 field.

The SXDF is a blank field with deep multi-wavelength observations from X-ray to radio. 
The AzTEC/ASTE observations covered $\sim$0.27~deg$^2$ of the central part of the SXDF with an rms noise level of $\sim$0.5--0.9~mJy (S.~Ikarashi et al. in preparation), which is about a factor of two deeper than the AzTEC/JCMT survey of this field \citep{aust10}. 
In total $\sim$200 sources ($\ge$3.5$\sigma$) are detected.

The SSA~22 field is thought to be a proto-cluster region since it has an overdensity of UV/optically selected galaxies such as Lyman-$\alpha$ emitters (LAEs) and Lyman-break galaxies (LBGs) at $z \sim 3.1$ \citep[e.g.,][]{stei98, stei00, haya04, matsuda05}. 
The SSA~22 field characterized by the overdensity is a good comparison field with other blank fields to see the relation between SMGs and other galaxy populations and the large-scale structure of the universe. 
The AzTEC/ASTE observations covered $\sim$0.28~deg$^2$ with an rms noise level of $\sim$0.6--1.2~mJy, and detected $\sim$100 sources ($\ge$3.5$\sigma$) \citep[][Tamura et al. in preparation]{tamu09}.

\begin{figure}
\begin{center}
\includegraphics[bb=100 50 360 302,clip,width=\linewidth]{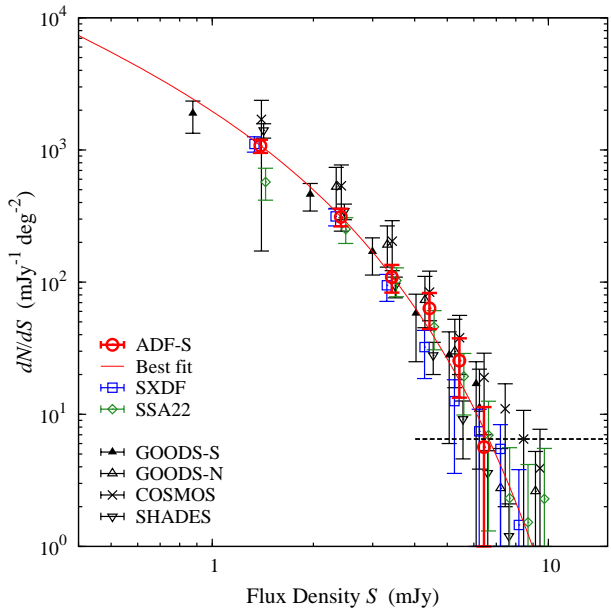}
\end{center}
\caption{
Differential number counts in the ADF-S, the SXDF, and the SSA~22 fields surveyed by AzTEC/ASTE. 
The counts are calculated for the 50\% coverage regions. 
The error bars represent the 68.3\% confidence intervals.
The red curve represents the best-fit Schechter functional form of the ADF-S: 
$dN/dS = N_{\rm 3 mJy}(S/{\rm 3\ mJy})^{-1} \exp(-(S-{\rm 3\ mJy})/S')$ with $N_{\rm 3 mJy} = 169 \pm 19$ and $S' = 1.48 \pm 0.20$. 
The number counts of previous 1.1~mm surveys are also shown: 
GOODS-N \citep{pere08}, COSMOS \citep{aust09}, SHADES \citep{aust10}, and GOODS-S \citep{scot10}. 
The horizontal dashed line represents the survey limit of the ADF-S, which Poisson deviations to zero sources per bin 32.7\%
. 
The bin centers of some number counts are shifted by $\pm$3\% for easier comparison. 
}
\label{fig:dif_counts}
\end{figure}

\begin{figure}
\begin{center}
\includegraphics[bb=100 50 360 302,width=\linewidth]{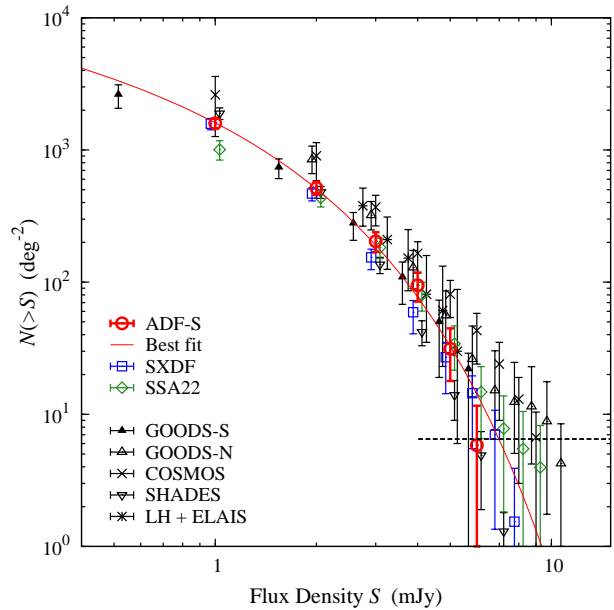}
\end{center}
\caption{
Cumulative number counts in the ADF-S, the SXDF, and the SSA~22 fields surveyed by AzTEC/ASTE. 
The counts are calculated for the 50\% coverage regions. 
The error bars indicate 68.3\% confidence intervals. 
The red curve represents the best-fit Schechter functional form of the ADF-S. 
Number counts of previous 1~mm surveys are also shown: 
the AzTEC/JCMT surveys of GOODS-N \citep{pere08}, COSMOS \citep{aust09}, and SHADES \citep{aust10}, the AzTEC/ASTE GOODS-S \citep{scot10}, 1.2~mm MAMBO surveys of the Lockman Hole and ELAIS N2 regions \citep{grev04}. 
The horizontal dashed line represents the survey limit of the ADF-S, which Poisson deviations to zero sources per bin 32.7\%. 
The bin centers of some number counts are shifted by $\pm$3\% for easier comparison. 
}
\label{fig:int_counts}
\end{figure}

The procedure and parameters in data reductions and extracting number counts of these two fields are the same as used in this paper. 
The differential and cumulative number counts are presented in Fig.~\ref{fig:dif_counts}, Fig.~\ref{fig:int_counts}, and Table \ref{tab:counts_sxdf_ssa22}. 
The best-fit parameters of the Schechter functional form in equation~(\ref{eq:schechter}) are shown in Table~\ref{tab:best-fit}.

\begin{table}
\caption{
Differential and cumulative number counts in the SXDF and the SSA~22 field. 
The flux bin centers for the differential counts (first column) are weighted by the prior. 
The errors are 68.3\% confidence intervals. 
}
\label{tab:counts_sxdf_ssa22}
\begin{center}
\begin{tabular}{@{}cccc@{}}
\hline\hline
Flux Density & Differential & Flux Density & Cumulative \\
(mJy) & (mJy$^{-1}$ deg$^{-2}$) & (mJy) & (deg$^{-2}$) \\
\hline
\multicolumn{4}{c}{SXDF} \\
\hline
1.38 & 1110$^{+148}_{-148}$ & 1.0 &  1578$^{+156}_{-158}$ \\
2.40 & 314 $^{+44 }_{-48 }$ & 2.0 &  468 $^{+50 }_{-56 }$ \\
3.41 & 95  $^{+20 }_{-23 }$ & 3.0 &  154 $^{+24 }_{-20 }$ \\
4.41 & 32  $^{+11 }_{-14 }$ & 4.0 &  59  $^{+13 }_{-19 }$ \\
5.42 & 13  $^{+5.7}_{-9.0}$ & 5.0 &  27  $^{+7.6}_{-13 }$ \\
6.42 & 7.4 $^{+3.5}_{-6.9}$ & 6.0 &  14  $^{+5.1}_{-8.9}$ \\
\hline
\multicolumn{4}{c}{SSA~22} \\
\hline
1.40 & 572$^{+152}_{-178}$ & 1.0 & 1006$^{+163}_{-191}$ \\
2.42 & 251$^{+52 }_{-60 }$ & 2.0 &  434$^{+59 }_{-69 }$ \\
3.43 & 103$^{+24 }_{-27 }$ & 3.0 &  182$^{+28 }_{-34 }$ \\
4.44 & 46 $^{+13 }_{-17 }$ & 4.0 &  80 $^{+16 }_{-21 }$ \\
5.44 & 19 $^{+7.7}_{-11 }$ & 5.0 &  34 $^{+8.8}_{-13 }$ \\
6.44 & 7.0$^{+3.3}_{-6.6}$ & 6.0 &  15 $^{+4.2}_{-7.0}$ \\
\hline
\end{tabular}
\end{center}
\end{table}

\begin{table}
\caption{
Best-fit parameters of parametric fits to differential number counts in the ADF-S, SXDF, and SSA~22 fields using the Schechter functional form in equation~(\ref{eq:schechter}). The errors are 1$\sigma$ uncertainty. 
}
\label{tab:best-fit}
\begin{center}
\begin{tabular}{@{}ccc@{}}
\hline\hline
Field & $N_{\rm 3 mJy}$ & $S'$ \\
      & (mJy$^{-1}$~deg$^{-2}$) & (mJy) \\
\hline
ADF-S & $169 \pm 19$ & $1.48 \pm 0.20$ \\
SXDF  & $144 \pm 17$ & $1.24 \pm 0.15$ \\
SSA~22& $132 \pm 18$ & $1.85 \pm 0.38$ \\
\hline
\end{tabular}
\end{center}
\end{table}

\subsection{Comparison among 1-mm Surveys}\label{sec:field_comparison}

In Figs~\ref{fig:dif_counts} and \ref{fig:int_counts}, we compare the number counts in the ADF-S, the SXDF, and the SSA~22 fields surveyed by AzTEC/ASTE with those of previous 1-mm survey; 
The AzTEC/JCMT surveys of GOODS-N \citep{pere08}, COSMOS \citep{aust09}, and SHADES (combined counts in the Lockman Hole and SXDF) \citep{aust10}, the AzTEC/ASTE survey of GOODS-S \citep{scot10}, and 1.2~mm MAMBO surveys of the Lockman Hole and ELAIS N2 \citep{grev04}.

The ADF-S and the SXDF provide the tightest constraints on the faint end of the number counts because of their depth and large survey areas. 
On the whole, the 1~mm counts of various surveys are consistent within errors. 
This is interesting since the SSA~22 field has overdensity of UV/optically-selected galaxies. 
It is possible that the overdensity of sources at $z=3.1$ traced by the UV/optical galaxies does not significantly change the SMG number counts given the large volume and redshift-space sampled by the mm-wavelength observations.
Compared to the ADF-S, the counts from GOODS-N and COSMOS are higher, while the counts from SHADES are lower. 
The overdensity of bright SMGs in AzTEC/COSMOS compared to other blank fields has been shown to be correlated with foreground structure at $z \la 1$ \citep{aust09}. 
Since the GOODS and the SHADES fields have no known biases, the diversity in the number counts likely arises from cosmic variance given the small areas of these surveys.

\subsection{Comparison with Models}\label{sec:model}

We compare the cumulative number counts with the models of \cite{take01a, take01b}, \cite{fran10}, and \cite{rowa09} which successfully reproduce the observed CIB and number counts at IR and submillimetre wavelengths. 
The models of \cite{fran10} and \cite{rowa09} are constructed to match observed 1.1~mm number counts of AzTEC surveys of COSMOS field \citep{aust09} and GOODS-N field \citep{pere08}, respectively.

\begin{figure}
\begin{center}
\includegraphics[bb=110 50 360 302,width=\linewidth]{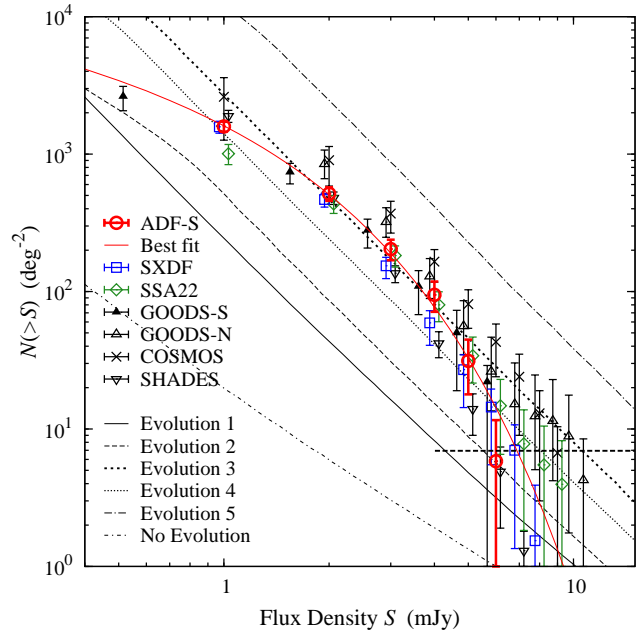}
\end{center}
\caption{
Comparison of 1.1~mm observed number counts with the models of \citet{take01a}. 
Descriptions of the models are in \S~\ref{sec:model}. 
}
\label{fig:int_counts_takeuchi}
\end{figure}

The model of \cite{take01a, take01b} consists of three components: (i) the FIR spectral energy distribution (SED) based on the \textit{IRAS} colour-luminosity relation at 60 $\mu$m and 100 $\mu$m; (ii) the local 60 $\mu$m luminosity function adopted from the \textit{IRAS} data; and (iii) galaxy evolution with redshift. 
The 60-$\mu$m luminosity of a galaxy is thus described as a function of redshift, assuming pure luminosity evolution: 
\begin{eqnarray}
L_{60}(z)=L_{60}(z=0)f(z), 
\end{eqnarray}
where $L_{60}$ is the luminosity at 60 $\mu$m and $L_{60}(z=0)$ represents the local 60 $\mu$m luminosity function.  
The form of $f(z)$ is a stepwise nonparametric function. 
\cite{take01a} assume three evolutionary scenarios \citep[see Figure~2 of][]{take01a} within the permitted range derived from the observed CIB and number counts at 15, 60, 90, 170, 450, and 850~$\mu$m:
(i) Evolution~1: $f(z)$ rises steeply between $z = 0-0.5$, reaching $f(z)=10$, is constant from $z = 0.5-2$, and decreases slowly between $z = 2-5$; 
(ii) Evolution~2: $f(z)$ quickly rises between $z = 0-0.5$, peaking with $f(z)=30$ between $z = 0.5-0.75$, and having a long plateau with $f(z)=10$ between $z = 0.75-5$; 
(iii) Evolution~3: $f(z)$ rises between $z = 0-1$, peaks with $f(z)=30$ between $z = 1-2$, and has a plateau with $f(z)=20$ between $z = 2-5$.
They adopt two additional evolution models which are made by modifying 
Evolution 1 to rise to $f(z)=10^{1.5}$ and $f(z)=10^{2.0}$ between $z = 1-2$, which we will refer to as ``Evolution 4'' and ``Evolution 5'', respectively. 
Figure~\ref{fig:int_counts_takeuchi} compares 1.1~mm observed number counts to the five models, along with a no-evolution model. 
The no-evolution, Evolution~1, and Evolution~2 models are lower than the observed counts, suggesting that significant luminosity evolution with $f(z) > 10$ is needed. 
Given that these five models fail to explain the observed counts, the models need to be modified. 
Possible ways to solve this discrepancy are 
(i) to change the functional form of the evolutionary scenarios, and 
(ii) to make the luminosity evolution dependent on luminosity. 
These issues are discussed in more detail in Takeuchi et al., in preparation.

The model of \cite{fran10} is constructed to reproduce the latest observed counts at 15, 24, 70, 350, 850, and 1100~$\mu$m, the redshift dependent luminosity functions at 15~$\mu$m, and the CIB. 
They assume both luminosity and number density evolution, and create number counts starting from the IRAS 12~$\mu$m luminosity function. 
The model population consists of four galaxy classes: 
non-evolving normal spirals, type-I AGNs, starburst galaxies of moderate luminosities (or LIRGs), and very luminous starburst galaxies (or ULIRGs). 
The four populations follow different evolution in luminosity and number density. 
The model of \cite{fran10} overestimates the ADF-S counts, while it is consistent with the counts of the AzTEC/COSMOS survey. 
This is not surprising, since the model is created to reproduce the 1.1~mm counts of the AzTEC/COSMOS, where a significant overdensity of sources has been reported \citep{aust09}.

\begin{figure}
\begin{center}
\includegraphics[bb=110 50 360 302,width=\linewidth]{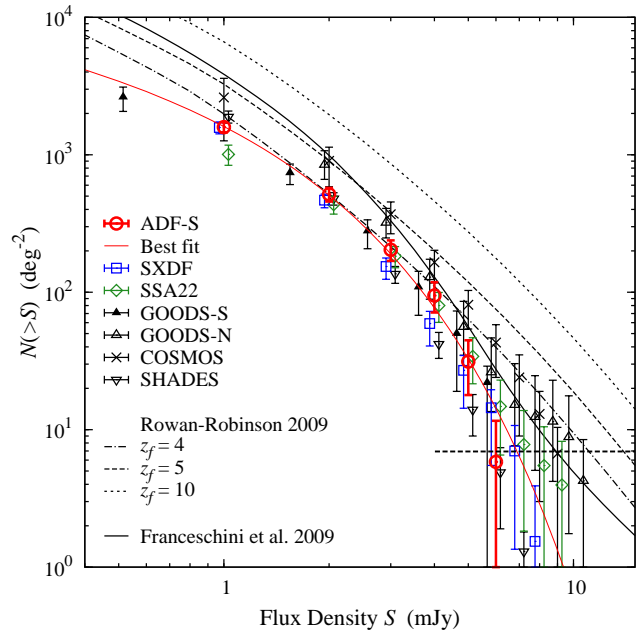}
\end{center}
\caption[Comparison of cumulative number counts with models]
{
Comparison of 1.1~mm observed number counts with the models of \citet{rowa09} and \cite{fran10}. 
Descriptions of the models are in \S~\ref{sec:model}. 
}
\label{fig:int_counts_model}
\end{figure}

The model of \cite{rowa09} is created by modifying the model of \cite{rowa01} to reproduce the latest observed counts, particularly at 24~$\mu$m. 
The model assumes pure luminosity evolution. 
The model consists of four spectral components: infrared cirrus, M82-like starburst, Arp 220-like starburst, and AGN dust torus. 
\cite{rowa09} creates three models with formation redshifts of $z_f = 4, 5$ and 10. 
Figure~\ref{fig:int_counts_model} shows that the model with $z_f = 4$ well describes the ADF-S counts down to the faint end, but at the bright end it overestimates the ADF-S counts. 

All of the models presented in this section do not match simultaneously the faint and bright end of observed counts, requiring the models to be modified.

\section{Contribution to Cosmic Infrared Background}\label{sec:cib}

We estimate the fraction of the CIB resolved by the ADF-S survey. 
The total deboosted flux density of the $\ge$3.5$\sigma$ sources in the 50\% coverage region is $1.9 \pm 0.03$~Jy~deg$^{-2}$. 
The expected 1.1~mm background as measured by the {\sl Cosmic Background Explorer} satellite is 18--24 Jy~deg$^{-2}$ \citep{puge96, fixs98}, therefore we have resolved about 7--10\% of the CIB into discrete sources. 
This is similar to the resolved fraction of the CIB reported by other 1~mm blank field surveys \citep{grev04, laur05, malo05, scot08, scot10}. 
This could be caused by the survey incompleteness due to the confusion noise and the fewer bright sources in the ADF-S, despite the deeper sensitivity compared to the other surveys. 

To estimate the total integrated flux density corrected for the survey incompleteness, and to include fainter sources below the detection threshold, we integrate the best-fit Schechter function of the ADF-S obtained in \S~\ref{sec:counts}. 
The integration of the best-fit function at $\ge$1~mJy, where the number counts are tightly constrained, is 2.9~Jy~deg$^{-2}$, which corresponds to 12--16\% of the CIB at 1.1~mm. 
This suggests that a large fraction of the CIB originates from submm-faint sources for which the number counts have not yet been constrained. 
Integration of the best-fit number counts extrapolating to lower fluxes (down to 0~mJy) results in a total flux density of 5.7~Jy~deg$^{-2}$, which is only 24--32\% of the CIB at 1.1~mm, suggesting that the faint-end slope of the actual number counts should be steeper than that of the present best-fit model. 
It is possible that a Schechter functional form is not appropriate for representing 1.1~mm number counts.

\section{Redshift Constraint}\label{sec:comparison}

\begin{figure}
\begin{center}
\includegraphics[width=\linewidth]{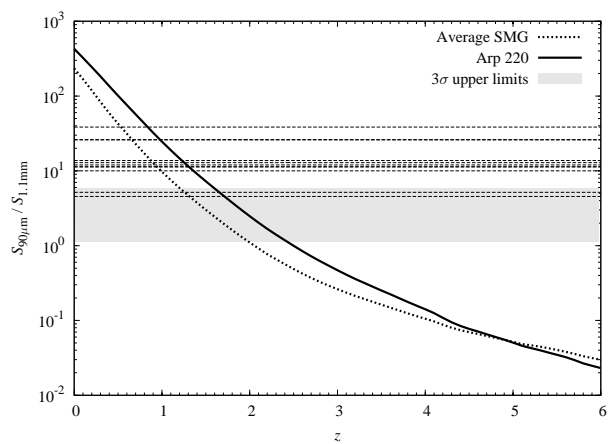}
\end{center}
\caption{
Observed flux ratios between 1.1~mm and 90~$\mu$m as a function of redshift. 
The solid and dashed curves represent the SED models of Arp~220 \citep{silv98} and the average SED of 76 SMGs with spectroscopic redshifts \citep{mich10}, respectively. 
The horizontal dotted lines indicate the flux ratios of AzTEC sources with 90~$\mu$m counterpart candidates. 
The shaded region represents the 3$\sigma$ upper limits of 90~$\mu$m for AzTEC sources without 90~$\mu$m detections. 
}
\label{fig:flux-ratio}
\end{figure}

In order to constrain redshifts of AzTEC sources, we compare AzTEC sources in the 30\% coverage region with the far-infrared images obtained by {\sl AKARI}/FIS (Shirahata et al. in prep.). 
The ADF-S is observed at four bands: 65, 90, 140, and 160 $\mu$m, with FWHMs of $37''$, $39''$, $58''$, and $61''$, respectively. 
The detection limits of the {\it AKARI} data are 46.5, 15.7, 183, 608 mJy (3$\sigma$) at 65, 90, 140, and 160 $\mu$m, respectively. 
We compare the AzTEC sources with the 90~$\mu$m source catalog, which is most sensitive and reliable among the four bands, and found only 11 AzTEC sources are within a $20''$ radius from the 90~$\mu$m sources. 
A detailed multiwavelength study of these sources will be made in a future paper.
We constrain the redshifts of the AzTEC sources using their flux ratios between 1.1~mm and 90~$\mu$m. 
Figure~\ref{fig:flux-ratio} shows the expected flux ratio as a function of redshift for two different SED models: 
Arp~220 \citep{silv98}, a typical ultra-luminous infrared galaxy (ULIRG), and the average SED of 76 SMGs with spectroscopic redshifts \citep{mich10}. 
The horizontal dotted lines represent the flux ratios of AzTEC sources with 90~$\mu$m counterpart candidates. 
The shaded region represents upper limits on the flux ratios of the remaining AzTEC sources from the 3$\sigma$ detection limit at 90~$\mu$m. 
The figure suggests that most of the AzTEC sources are likely to be at $z \gsim 1.5$. 

At z $>$ 1, the flux density of SMGs is nearly redshift independent and thus is proportional to the IR luminosity. 
By scaling the IR luminosities of the SED models, we estimate IR luminosities of the AzTEC sources to be $\sim$3--14 $\times 10^{12}$~$\LO$. 
If the emission is powered solely by star formation activity, the inferred SFRs are $\sim$500--2400~$\MO$~yr$^{-1}$ \citep{kenn98}.

\section{Cosmic Star Formation History Traced by 1.1~mm Sources}\label{sec:sfh}

\subsection{Star-formation Rate Density}\label{sec:sfrd}

From UV/optical surveys, the total SFR per unit comoving volume (SFR density) is observed to increase with redshift from $z \sim 7$ to $z \sim 3$, peak at $z \sim 3$--1, and decline steadily from $z = 1$--0 \citep[e.g.,][]{mada96, lill96, stei99, giav04, bouw10}. 
It is suggested that SFRs per unit comoving volume (SFR density) increases with redshift from $z \sim 7$ to $z \sim 3$, peaks at $z \sim 1$--3, and decreases from $z \sim 1$. 
However, the SFR density derived by UV/optically selected galaxies has large uncertainty due to the extinction by dust. 
It is also possible that dusty galaxies are missed entirely by these previous studies. 
Comparatively speaking, millimetre and submillimetre wavelengths have a great advantage in tracing dusty starburst galaxies at high redshifts. 
Previous submillimetre surveys suggested that SMGs contribute significantly ($\gsim$10--20\%) to the cosmic SFR density at $z \sim 2$--3 \citep[e.g.,][]{hugh98, chap05, aret07, dye08, ward10}. 

We estimate the SFR density contributed by 1.1~mm sources using the best-fit number counts in the ADF-S derived in \S~\ref{sec:counts}. 
FIR luminosities of the 1.1~mm sources are calculated by assuming the SED models of Arp~220 \citep{silv98} and the average SED of SMGs \citep{mich10}, and SFRs are derived from FIR luminosity using the equation of \cite{kenn98}. 
The largest uncertainty comes from the lack of redshift information. 
Since the redshifts of the 1.1~mm sources are not known, we assume redshift distributions based on previous studies. 
The largest spectroscopic sample of SMGs obtained by \cite{chap05} has a median redshift of $z = 2.2$ with an interquartile range of 1.7--2.8, and the redshift distribution is well fitted by Gaussian. 
\cite{pope06} found a median redshift of $z = 2.2$ with an interquartile range of 1.4--2.6 using spectroscopic and photometric redshifts of SMGs in the Hubble Deep Field-North.  
\cite{aret07} estimated photometric redshifts of SHADES sources and found a median of $z = 2.4$ with an interquartile range of $z = 1.8$--3.1, and that redshift distribution has a near-Gaussian form. 
\cite{chap09} derive a higher median redshift of $z = 2.7$ using spectroscopic and photometric redshifts of 1.1~mm sources detected in the AzTEC/JCMT GOODS-N survey.

Based on these measurements, we adopt Gaussian functional forms of redshift distributions with central redshifts of $z_c =2.2, 2.4$, and 2.7. 
We assume $\sigma_z = 0.5$ and $\sigma_z = 1.0$ for narrow and broad redshift distributions, respectively. 
These redshift distributions are consistent with the fact that the most of the AzTEC sources are at $z \gsim 1.5$ (see \S~\ref{sec:comparison}). 
We calculate the total flux by integrating the differential number counts at $\ge$1~mJy and distribute the total flux following the assumed redshift distributions. 
If we integrate the number counts down to 0.1~mJy, the total flux density would increase by about a factor of two.

\begin{figure}
\begin{center}
\includegraphics[width=\linewidth]{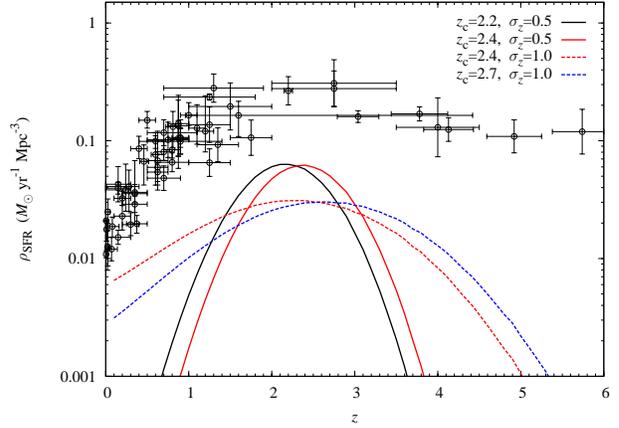}
\end{center}
\caption{
Star-formation rate density estimated from the best-fit number counts in the ADF-S. 
Four redshift distributions are assumed. 
The SFR densities are calculated by integrating the number counts at $\ge$1~mJy and using the SED model of Arp~220. 
The results from previous UV-optical observations are plotted for comparison \citep[compiled data are taken from][and references therein]{hopk04}. 
}
\label{fig:sfrd}
\end{figure}

The derived SFR densities using the Arp~220 SED model are shown in Figure~\ref{fig:sfrd} and the average values in redshift bins are presented in Table~\ref{tab:sfrd}. 
The results derived from the two assumed SED models are consistent within 30\%. 
Although the derived SFR densities largely depend on the assumed redshift distribution, they are within the range of those derived in previous studies with SMGs \citep{hugh98, chap05, aret07, dye08, ward10}. 
In Figure~\ref{fig:sfrd}, extinction-corrected SFR densities derived from previous UV/optical observations are also shown for comparison \citep{hopk04}. 
The SFR densities of 1.1~mm sources are lower by about a factor of 5--10 at $z \sim 2$--3 compared to that of the UV/optically-selected galaxies.

In \S~\ref{sec:cib}, we found that integrating the AzTEC/ADF-S number counts at $\ge$1~mJy accounts for 12--16\% of CIB at 1.1~mm. 
If we assume that the rest of the CIB comes from fainter sources, 
the total SFR density contributed by 1.1~mm sources including $<$1~mJy sources would increase by about a factor of 6--8, which is comparable to or higher than that of the UV/optically-selected galaxies at $z \sim 2$--3. 
We note that in this case the faint 1.1~mm sources and UV/optical sources can overlap. 
The large contribution of dusty galaxies to the SFR density is suggested by \cite{goto10} based on 8~$\mu$m and 12~$\mu$m observations. 
They found that the dust-obscured SFR density at $z \sim 2$ is $\sim$0.5~$\MO$~yr$^{-1}$~Mpc$^{-3}$, which is consistent with the SFR density of 1.1~mm sources including the fainter ($<$1~mJy) population sources.

\begin{table}
\centering
\caption{
Comoving star formation rate density averaged in redshift bins estimated from the best-fit number counts in the ADF-S. 
Four redshift distributions are assumed. 
The SFR densities are calculated by integrating the number counts at $\ge$1~mJy and using the SED model of Arp~220. 
}
\label{tab:sfrd}
\begin{tabular}{@{}ccccc@{}}
\hline\hline
($z_c, \sigma_z$) & \multicolumn{4}{c}{Comoving Star-formation Rate Density} \\
 & \multicolumn{4}{c}{($\MO$~yr$^{-1}$~Mpc$^{-3}$)} \\
\hline
 & $z=1$--2 & $z=2$--3 & $z=3$--4 & $z=4$--5 \\
\hline
(2.2, 0.5) & $3.0 \times 10^{-2}$ & $4.6 \times 10^{-2}$ & $4.1 \times 10^{-3}$ & $1.2\times 10^{-5}$ \\
(2.4, 0.5) & $1.9 \times 10^{-2}$ & $5.1 \times 10^{-2}$ & $8.5 \times 10^{-3}$ & $5.2\times 10^{-5}$ \\
(2.4, 1.0) & $2.3 \times 10^{-2}$ & $2.9 \times 10^{-2}$ & $1.6 \times 10^{-2}$ & $3.8\times 10^{-3}$ \\
(2.7, 1.0) & $1.7 \times 10^{-2}$ & $2.9 \times 10^{-2}$ & $2.1 \times 10^{-2}$ & $6.5\times 10^{-3}$ \\
\hline
\end{tabular}
\end{table}

\subsection{Stellar Mass Density}\label{sec:smd}

We estimate the fraction of the stellar mass in the present-day universe produced by 1.1~mm sources with $\ge$1~mJy by integrating the SFR density derived in the previous section. 
The present-day stellar mass density is estimated from local luminosity functions \citep[e.g.,][]{cole01, bell03, kaji09}. 
\cite{cole01} derived the present-day stellar mass density of ($5.6 \pm 0.8$)~$\times 10^8~\MO$~Mpc$^{-3}$ assuming a \cite{salp55} initial mass function. 
The time integration of the SFR densities at $z \ge 1$ for the four assumed redshift distributions (($z_c, \sigma_z$) =(2.2, 0.5), (2.4, 0.5), (2.4, 1.0), and (2.7, 1.0)) yields 
$1.2 \times 10^8~\MO$~Mpc$^{-3}$, 
$1.0 \times 10^8~\MO$~Mpc$^{-3}$, 
$1.0 \times 10^8~\MO$~Mpc$^{-3}$, 
and $0.90 \times 10^8~\MO$~Mpc$^{-3}$, respectively. 
This corresponds to $\sim$20\% of the present-day stellar mass density. 
This is an upper limit since materials forming massive stars returned to the interstellar medium via stellar winds and supernovae explosions. 
The fraction of stellar mass returned to the ISM, called as the recycled fraction, is estimated in semi-analytical models \citep[e.g.,][]{cole00, baug05, lace10, gonz10}, and it depends on the IMF: 
0.41 for the \cite{kenn83} IMF and 0.91 for a top-heavy IMF \cite{cole00, lace10}. 
If we assume the recycled fraction of 0.41 and 0.91, the fraction of the stellar mass in the present-day universe produced by 1.1~mm sources with $\ge$1~mJy decrease to $\sim$10\% and a few percent, respectively.

\section{Summary}\label{sec:summary}

We performed a 1.1~mm deep survey of the AKARI Deep Field South (ADF-S) with AzTEC mounted on the ASTE, obtaining one of the deepest and widest maps at millimetre wavelengths. 
The 30\% and 50\% coverage regions have areas of 909 and 709~arcmin$^2$, and noise levels of 0.32--0.71~mJy and 0.32--0.55~mJy, respectively. 
We detected 198 previously unknown millimetre-bright sources with 3.5--15.6$\sigma$ in the 30\% coverage region, providing the largest 1.1~mm source catalog from a contiguous region.

We constructed differential and cumulative number counts in the ADF-S, the SXDF, and the SSA~22 field which probe fainter flux densities (down to 1~ mJy) compared to previous surveys except for the AzTEC/ASTE GOODS-S survey. 
On the whole, the 1~mm counts of various surveys are consistent within errors. 
We compare the number counts with the luminosity evolution models of \cite{take01a}, \cite{fran10}, and \cite{rowa09}. 
Comparison with the \cite{take01a} model suggests that a luminosity evolution with a factor $>$10 is needed to explain the observed number counts. 
The observed number counts favor the model of \cite{rowa09} with $z_f = 4$, but none of these models simultaneously match both the bright and faint end of the number counts from 1--10~mJy.

In the ADF-S survey, we resolve about 7--10\% of the CIB at 1.1~mm into discrete sources. 
The integration of the best-fit number counts in the ADF-S down to 1~mJy reaches 12--16\% of the CIB. 
This suggests that the large fraction of the CIB at 1.1~mm originates from faint sources ($S_{\rm 1.1mm} < 1$~mJy) for which the number counts have not yet been constrained. 
The integration of the best-fit number counts extrapolating to 0~mJy accounts for only 24--32\% of the CIB, suggesting that the faint-end slope of the number counts is steeper than that given by our best-fit model.

The redshifts of the AzTEC sources are constrained from their flux ratios between 1.1~mm and 90~$\mu$m from the {\sl AKARI}/FIS. 
Most of the AzTEC sources are not detected at 90~$\mu$m, suggesting that they are likely to be at $z \gsim 1.5$. 
Assuming $z \gsim 1$, the inferred IR luminosities of the AzTEC sources are $\sim$(3--14)$\times 10^{12}~\LO$, and their SFRs inferred from the IR luminosities are $\sim$500--2400~$\MO$~yr$^{-1}$.

We derived the cosmic SFR density contributed by 1.1~mm sources using the best-fit model to the differential number counts. 
Although the derived SFR density largely depends on the assumed redshift distribution, our estimates are within the range of those derived in previous studies with SMGs. 
The SFR density of 1.1~mm sources with $\ge$1~mJy at $z \sim 2$--3 is lower by about a factor of 5--10 than those of UV/optically-selected galaxies. 
If we consider the fact that the contribution of 1.1~mm sources with $\ge$1~mJy to the CIB at 1.1~mm is 12--16\%, the SFR density of 1.1~mm sources including those fainter than $S_{\rm 1.1mm} < 1$~mJy would become comparable to or higher than those of UV/optically-selected galaxies. 
The fraction of the present-day stellar mass of the universe produced by 1.1~mm sources with $\ge$1~mJy at $z \ge 1$ is $\sim$20\%, calculated by the time integration of the SFR density. 
If we consider the recycled fraction of 0.41 and 0.91, the fraction of stellar mass produced by 1.1~mm sources becomes $\sim$10\% and a few percent, respectively.

\section*{Acknowledgments}
We are grateful to K.~Shimasaku, H.~Murakami, S.~Okumura, N~.Yoshida, T.~Kodama, and the anonymous referee for helpful comments and suggestions. 
We would like to thank Alberto Franceschini for providing the data of his number counts models.  
BH and YT were financially supported by a Research Fellowship from the JSPS for Young Scientists. 
TTT was supported by Program for Improvement of Research Environment for Young Researchers from Special Coordination Funds for Promoting Science and Technology, and the Grant-in-Aid for the Scientific Research Fund (20740105) commissioned by the Ministry of Education, Culture, Sports, Science and Technology (MEXT) of Japan. 
TTT was partially supported from the Grand-in-Aid for the Global COE Program ``Quest for Fundamental Principles in the Universe: from Particles to the Solar System and the Cosmos'' from the MEXT. 
This study was supported by the MEXT Grant-in-Aid for Specially Promoted Research (No.~20001003).
AzTEC/ASTE observations were partly supported by was supported by the Grant-in-Aid for the Scientific Research from the Japan Society for the promotion of Science (No. 19403005)
The ASTE project is driven by Nobeyama Radio Observatory (NRO), a branch of National Astronomical Observatory of Japan (NAOJ), in collaboration with University of Chile, and Japanese institutes including University of Tokyo, Nagoya University, Osaka Prefecture University, Ibaraki University, and Hokkaido University. 
Observations with ASTE were in part carried out remotely from Japan by using NTT's GEMnet2 and its partnet R\&E (Research and Education) networks, which are based on AccessNova collaboration of University of Chile, NTT Laboratories, and NAOJ.
AzTEC analysis performed at UMass is supported in part by NSF grant 0907952.


\end{document}